\documentclass[12pt]{article}
\usepackage{amsmath}
\usepackage{graphicx}
\usepackage[round,authoryear]{natbib}
\usepackage{url} 

\usepackage{amssymb}

\usepackage{amsmath}
\usepackage{amsfonts,amsmath}
\usepackage{amsthm}
\usepackage{mathtools}
\usepackage[english]{babel}
\usepackage{amssymb}
\usepackage{graphicx}
\usepackage{algorithm,algpseudocode}
\usepackage{color}
\usepackage{lipsum}
\usepackage{mwe}
\usepackage{comment}
\usepackage{MnSymbol}
\excludecomment{confidential}
\newtheorem{theorem}{Theorem}[section]

\usepackage{bm}
\newtheorem{proposition}{Proposition}
\usepackage{caption}
\usepackage{float}
\usepackage{placeins}
\usepackage{subcaption}

\newcommand{\blind}{0}
\DeclareMathOperator*{\argmax}{arg\,max}

\allowdisplaybreaks

\begin{document}

\bibliographystyle{apalike}

\def\spacingset#1{\renewcommand{\baselinestretch}%
{#1}\small\normalsize} \spacingset{1}


\if0\blind
{
  \title{\bf EM Approaches to Nonparametric Estimation for Mixture of Linear Regressions}
  \author{Andrew Welbaum\hspace{.2cm}\\
    Department of Statistics, George Mason University\\
    and \\
    Wanli Qiao \\
    Department of Statistics, George Mason University}
  \maketitle
} \fi

\if1\blind
{
  \bigskip
  \bigskip
  \bigskip
  \begin{center}
    {\LARGE\bf Title}
\end{center}
  \medskip
} \fi

\bigskip
\begin{abstract}
In a mixture of linear regression model, the regression coefficients are treated as random vectors that may follow either a continuous or discrete distribution. We propose two Expectation-Maximization (EM) algorithms to estimate this prior distribution. The first algorithm solves a kernelized version of the nonparametric maximum likelihood estimation (NPMLE). This method not only recovers continuous prior distributions but also accurately estimates the number of clusters when the prior is discrete. The second algorithm, designed to approximate the NPMLE, targets prior distributions with a density. It also performs well for discrete priors when combined with a post-processing step. We study the convergence properties of both algorithms and demonstrate their effectiveness through simulations and applications to real datasets.
\end{abstract}

\noindent%
\vfill

\newpage
\spacingset{1.75} 
\section{Introduction}
\label{sec:intro}

The mixture of linear regression model extends the multiple linear regression by allowing the coefficient vector to have a continuous or discrete prior distribution, and has a wide number of applications when the response and covariates can have individualized or clustered linear relation, including market segmentation \citep{wedel2000market}, medical studies \citep{schlattmann2009medical}, educational research \citep{ding2006using}, and various industry and economic studies, such as housing construction \citep{quandt1972new}, wage prediction \citep{quandt1978estimating}, and climate-economic growth relationships \citep{yao2015mixtures}.  The model we consider in this article is as follows: There are $n$ independent observations $(x_1,y_1),\cdots,(x_n,y_n)\in\mathbb{R}^d\times\mathbb{R}$ generated from the model 
\begin{align}
\label{crmodel}
y_i = x_i^\top \beta_i + \sigma z_i,
\end{align}
where $\beta_1,\cdots,\beta_n,z_1,\cdots,z_n$ are independent of $\beta_1,\cdots,\beta_n \overset{iid}{\sim} G^*$ and $z_1,\cdots,z_n \overset{iid}{\sim}N(0,1)$, and $\sigma>0$. Here, $G^*$ is a probability distribution on $\mathbb{R}^d$, which may be discrete or continuous. 

We will assume that $\sigma$ is known, but $G^*$, the true distribution, is not known and needs to be estimated. In practice, an unknown $\sigma$ can be estimated using cross-validation.

The estimation of $G^*$ can be further used to make statistical inference for individual $\beta_i$ through a plug-in approach. 
The posterior distribution of $\beta_i$ given $(x_i,y_i)$ with respect to a prior distribution $G$ (which is $G^*$ if the prior is correctly identified) is 
\begin{align}
\label{posterior}
P(\beta_i \in A\; | \; x_i,y_i, G ) = \frac{\int_A \phi_\sigma\left(y_i - x_i^\top \beta\right) d G(\beta) }{\int_{\mathbb{R}^d}  \phi_\sigma\left( y_i - x_i^\top \beta\right) d G(\beta) },
\end{align}
for any measurable set $A\subset \mathbb{R}^d$,
where $\phi$ is the standard normal density, and $\phi_\sigma(\cdot)=(1/\sigma)\phi(\cdot/\sigma)$. The posterior mean of $\beta_i$ given $(x_i,y_i)$ is 
\begin{align}
\mathbb{E}(\beta_i \; | \; x_i,y_i, G ) = \frac{ \int_{\mathbb{R}^d}  \phi_\sigma\left( y_i - x_i^\top \beta\right) \beta d G(\beta)}{\int_{\mathbb{R}^d}  \phi_\sigma\left( y_i - x_i^\top \beta\right) d G(\beta) }.
\end{align} 
We can then obtain a point estimate of $\beta_i$ if we replace $G$ by an estimator of $G^*$ in $\mathbb{E}(\beta_i \; | \; x_i,y_i, G )$.

When $G^*$ is discrete, an additional interesting task is to cluster $(x_i,y_i)$'s based on this estimation. Suppose $G^*(\beta) = \sum_{j=1}^K \pi_j \delta_{\beta_j}(\beta)$, where $K$ is unknown, $\pi_j$ represents the weight of the component $j$ such that $\sum_{k=1}^{K} \pi_{k}=1$ and $\pi_k> 0$, and $\delta_{\beta_j}$ is the Dirac delta with point mass at $\beta_j$. The posterior distribution in \eqref{posterior} when $G=G^*$ takes the form of
\begin{align}
\label{discretedist}
\sum_{j=1}^K\left[\dfrac{\pi_j\phi_\sigma\left( y_i - x_i^\top \beta_j \right)}{\sum_{k=1}^{K}\pi_k\phi_\sigma\left( y_i - x_i^\top \beta_k\right)}\right]\delta_{\beta_j}(\beta).
\end{align}
In an oracle setting, if we knew the true prior distribution $G^*=G$, we could use this to identify cluster membership of the points $(x_i,y_i)$'s by assigning $(x_i,y_i)$ to the component $j$ with the highest posterior probability, i.e.,
\begin{align}
\label{posteriordiscrete}
\argmax_{j\in\{1,\dots,K\}} \dfrac{\pi_j\phi_\sigma\left( y_i - x_i^\top \beta_j\right)}{\sum_{k=1}^{K}\pi_k\phi_\sigma\left( y_i - x_i^\top \beta_k \right)}.
\end{align}
If the estimator of $G^*$ is in the form of $\sum_{j=1}^{\hat K} \hat\pi_j \delta_{\hat\beta_j}$, the above clustering rule is updated by replacing $\{(\pi_k,\beta_k): k=1,\dots,K\}$ with $\{(\hat\pi_k,\hat\beta_k): k=1,\dots,\hat K\}$.

A nonparametric approach to estimating $G^*$ is nonparametric maximum likelihood estimation (NPMLE), which does not assume a parametric form of $G^*$. The NPMLE for mixture models was first described by \citet{kiefer1956consistency}. Some recent NPMLE-related works include \citet{ jiang2009general,koenker2014convex,dicker2016high,jagabathula2020conditional,polyanskiy2020self,saha2020nonparametric,deb2021two,gu2022nonparametric}; and \citet{jiang2025nonparametric}. Of note, \citet{jiang2009general} introduce a general maximum likelihood empirical Bayesian method for estimatng a vector of means in a mixture of Gaussians problem, assuming independent and identically distributed errors. Their method aims to find the MLE of the distribution of the unknown parameters with a discrete set of support points using an EM algorithm. \citet{koenker2014convex} introduce an interior point method that estimates the nonparametric MLE of the Gaussian mixture problem. Also,  \citet{jiang2025nonparametric} describe an NPMLE approach to the mixture of linear regression problem using a method called Conditional Gradient Method (CGM). 

We now describe the NPMLE to estimate $G^*$. The log-likelihood function for model \eqref{crmodel} with a prior distribution $G$ given $(x_1,y_1),\cdots,(x_n,y_n)$ is 
\begin{align}
\label{incomplete-log}
G \mapsto L(G): = \sum_{i=1}^n \log f_{x_i}^G(y_i),
\end{align}
where
\begin{align}
\label{fy}
f_{x_i}^G(y_i) = \int \phi_\sigma\left(y_i - x_i^\top \beta\right) dG(\beta), \;\; i=1,\cdots,n.
\end{align}
We seek a maximizer of $L$ over the set $\mathcal{G}$ of probability distributions $G$ supported on $\mathcal{X}\subset\mathbb{R}^d$. That is, the NPMLE problem can be stated as 
\begin{align}
\label{NPMLE}
\hat G \in \argmax_{G\in\mathcal{G}} L(G).
\end{align}

If $\mathcal{X}$ is compact or satisfies a regularity condition met by $\mathbb{R}^d$, \citep{jiang2025nonparametric} establishes the existence of an NPMLE $\hat{G}$ that is a probability measure supported on at most $n$ points in $\mathcal{X}$ and is a consistent estimator of $G^*$ in terms of the L\'{e}vy-Prokhorov metric. 

If $\mathcal{G}$ is the class of mixture of $K$ probability distributions where $K$ is known, that is, 
\begin{align*}
\mathcal{G}=\mathcal{G}_K:=\left\{G = \sum_{j=1}^K \pi_j \delta_{\beta_j}: \;\; \sum_{j=1}^K \pi_{j} = 1, \;\pi_j> 0, \;\beta_j \in\mathbb{R}^d\right\},
\end{align*}
then the model becomes a finite mixture of linear regressions, and this problem can be solved by the regular EM algorithm for a known number of components \citep{leisch2004flexmix}. The modern EM algorithm was formalized in \citet{dempster1977maximum}. \citet{desarbo1988maximum} introduced an EM algorithm for the cluster of linear regression problem with $K$ components, which extended previous work on estimating two-component models such as in \citet{quandt1972new,david1974maximum}; and \citet{quandt1978estimating}.

\begin{figure}[H]
    \centering
    \begin{subfigure}[t]{0.45\textwidth}
        \centering
        \includegraphics[width=0.9\linewidth]{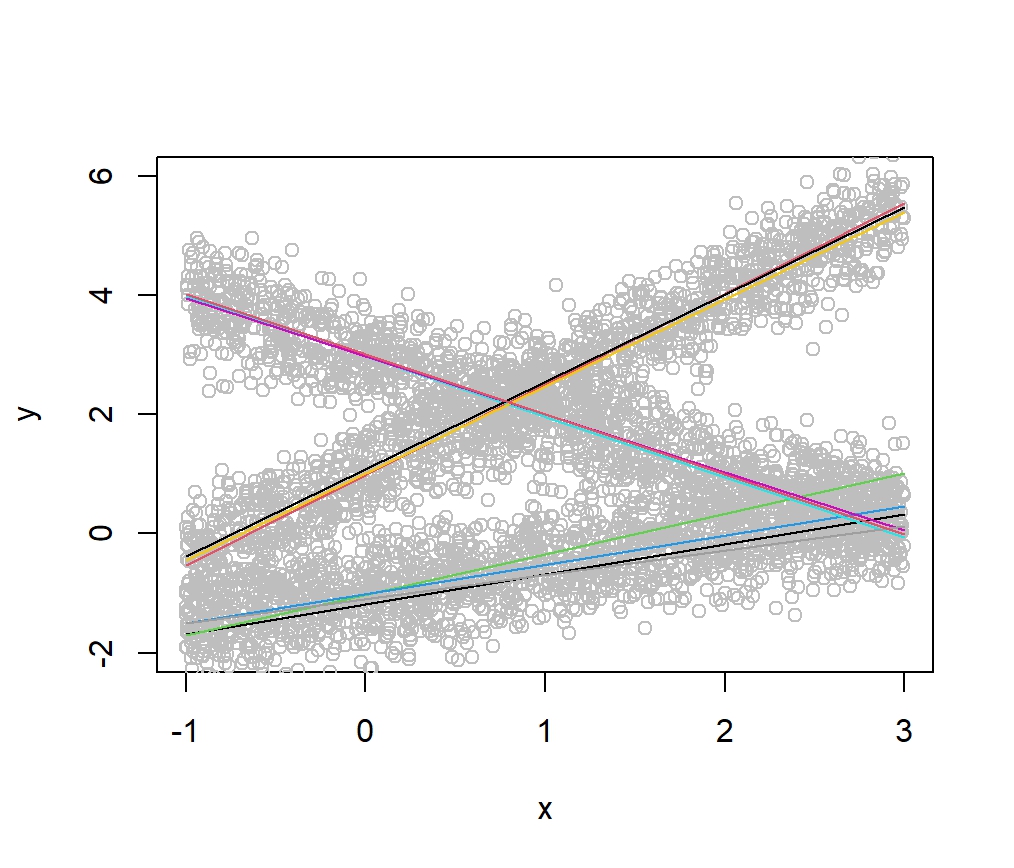}
    \end{subfigure}%
    \hspace{0.05\textwidth}
    \begin{subfigure}[t]{0.45\textwidth}
        \centering
        \includegraphics[width=0.9\linewidth]{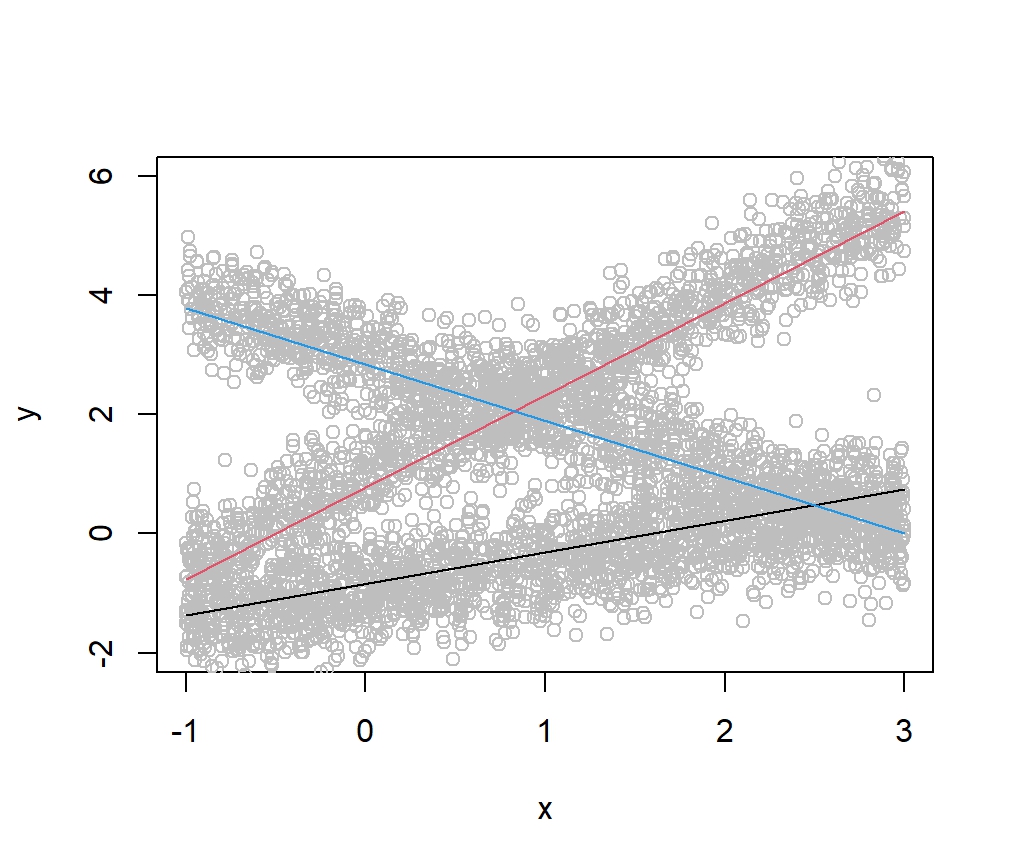}
    \end{subfigure}
    \caption{Estimated regression lines for a mixture of linear regression model with three true components: $y=3-x$, $y=1+1.5x$, and $y=-1+.5x$ with true weights (.3, .3, and .4), respectively, and noise with $\sigma=.5$. Gray points are $(x_i,y_i)$'s of size $n=5000$. Left panel: NPMLE solution estimated using the CGM method in \citet{jiang2025nonparametric}. There are 10 estimated regression lines with weights ranging from 0.012 to 0.343, although three clusters appear to be formed with aggregated weights 0.310, 0.296, and 0.393. Right panel: NPKMLE solution estimated by an EM algorithm developed in this paper. There are exactly 3 estimated regression lines with weights 0.314, 0.299 and 0.386, respectively.}
    \label{fig:all_methods_intro}
\end{figure}

In the classic EM algorithm, the number of components, if it is unknown, must be estimated prior to initiating the algorithm. Some approaches to estimating this number include methods based on the incomplete and complete log-likelihoods, Fischer information matrix, Bayesian criteria, and information theory (e.g., AIC) \citep{hawkins2001determining, melnykov2012initializing}. On the other hand, methods based on NPMLE such as the approach in \citet{jiang2025nonparametric} do not appear to be able to accurately detect the true number of components in the mixture model. Although the consistency for NPMLE has been established in \citet{jiang2025nonparametric}, in practice, a true cluster may correspond to a few small clusters detected by NPMLE, where the sum of the estimated weights approximate the true weight of that single cluster (see Figure \ref{fig:all_methods_intro}).
In our approach, instead of maximizing \eqref{NPMLE} without imposing a form restriction on $\mathcal{G}$, we restrict the set $\mathcal{G}$ to $\mathcal{G}_{\text{kde}}$, which is the set of kernel density estimators with a given kernel function and a bandwidth based on $n$ points in $\mathbb{R}^d$ and solve
\begin{align}
\label{newargmax}
\hat G_{\text{kde}} \in \argmax_{G\in\mathcal{G}_{\text{kde}}} L(G).
\end{align}
We call this the nonparametric kernel maximum likelihood estimator (NPKMLE). The optimization is reduced to finding the $n$ points that define the KDE solution since the kernel function and the bandwidth are given. 

We develop an EM algorithm that aims to find the solution to the above optimization problem. The algorithm is able to estimate a mixture of linear regression model with a true prior distribution that is discrete or continuous, as opposed to the classical EM-algorithm's restriction to estimating the parameters of a mixture of regression model with a finite number of components. In the case of a discrete prior distribution, our algorithm does not require a prior estimate of the number of components, and is able to determine the true number of components automatically (see Figure \ref{fig:all_methods_intro}). Hence the output of our algorithm can be used for clustering in the finite mixture of linear regression model. 

If the prior distribution $G^*$ is known to have a density, we also develop another EM algorithm that converges to NPMLE. This EM algorithm falls short when the density does not exist. However, specific post-processing steps tailored for the underlying structure of $G^*$ can remediate the issue, for example, when $G^*$ is supported on a finite set or has a low dimensional manifold as it support. 

This article is structured as follows: In Section \ref{sec:our approach}, we present the mixture of linear regression problem and discuss our approaches to solving it. We introduce an EM algorithm that can be used when the true distribution $G^*$ has a density, as well as a different EM algorithm when we drop this assumption. We also discuss the algorithms' properties. In Section \ref{sec:Simulations and Applications}, we demonstrate the performance of our algorithms using simulations and show their utility in real-world applications. All the proofs are provided in the Appendix.

\section{Methods}
\label{sec:our approach}

In the discussion below, we distinguish two cases between $G^*$ has a density and $G^*$ does not necessarily have a density (e.g., when $G^*$ is discrete with an unknown number of components), and describe an EM algorithm for each case. 

Our starting point is the NPMLE. Note that $L(G)$ in \eqref{incomplete-log} is the ``incomplete-data'' log-likelihood function (meaning that the information of $\beta_i$ for each pair $(x_i,y_i)$ is not available). If $G$ has a density $g$, then, abusing the notation $L$, which was used for $L(G)$, the log-likelihood can be written as 
\begin{align}
\label{incomplete-log2}
g \mapsto L(g): = \sum_{i=1}^n \log\left[ \int \phi_\sigma\left( y_i - x_i^\top \beta\right) g(\beta)d\beta \right].
\end{align}
The corresponding ``complete-data'' log-likelihood function is 
\begin{align}
\label{complete-log}
L(G| \bm{x,y,\beta}) = \sum_{i=1}^n \log \left[\phi_\sigma\left( y_i - x_i^\top \beta_i\right) g(\beta_i) \right],
\end{align}
where $\bm{x}=(x_1,\cdots,x_n)$, $\bm{y}=(y_1,\cdots,y_n)$, and $\bm{\beta}=\{\beta_1,\cdots,\beta_n\}$. Here $g$ should be understood as the density of $G$ if it exists; If $G$ is discrete such that $G = \sum_{j=1}^K \pi_j \delta_{\beta_j}$, then $g(\beta)=\sum_{j=1}^K \pi_j\mathbf{1}_{\beta_j}(\beta)$, where $\mathbf{1}_{\beta_j}$ is the indicator function at $\beta_j$.

\subsection{EM-NPMLE algorithm: when $G^*$ has a density function} 
\label{subsecsec:EM alg G density}

We first assume $G^*$ is a continuous distribution for which a density function $g^*$ exists. In this case, the posterior density of $\beta_i$ given $(x_i,y_i)$ with respect to $g^*$ exists and is 
\begin{align}
\label{fi}
f_i (\beta\; | \; x_i,y_i, g^* ) = \frac{ \phi_\sigma\left(y_i - x_i^\top \beta\right) g^*(\beta) }{\int_{\mathbb{R}^d}  \phi_\sigma\left( y_i - x_i^\top \beta\right) g^*(\beta)d\beta }.
\end{align}  
We provide an iterative algorithm to approximate $g^*$ in what follows. With an initial estimate of $g^*$, we can obtain the posterior density of $\beta_i$ given $(x_i,y_i)$ with respect to this initial estimate. Then it is a natural idea to use the average of the posterior densities across all $\beta_i$ to update the estimate of $g^*$, and the process is continued iteratively until convergence. More formally, with an initialization density $g^{(0)}$ for $\beta_i$'s, for $t=0,1,2,\cdots,$ the algorithm iterates between
\begin{align}
    f_i^{(t+1)}(\beta) \equiv f_i (\beta\; | \; x_i,y_i, g^{(t)} ) = \frac{ \phi_\sigma\left( y_i - x_i^\top \beta \right) g^{(t)}(\beta) }{\int_{\mathbb{R}^d}  \phi_\sigma\left( y_i - x_i^\top \beta \right) g^{(t)}(\beta)d\beta  },\; i=1,\cdots,n,
\end{align}
and
\begin{align}
\label{gtplus1old}
g^{(t+1)} = \frac{1}{n} \sum_{i=1}^n f_i^{(t+1)}.
\end{align}

It turns out that this simple but elegant algorithm can be interpreted as an EM algorithm, and it is inspired by a similar algorithm in the setting of mixture of distributions~\citep{vardi1985statistical,laird1991smoothing,vardi1993image,chung2015convergence,chae2018convergence}. We will call this an EM-NPMLE algorithm. To see this is an EM algorithm, taking the expectation of $L(g|\bm{x},\bm{y},\bm{\beta})$ (we again abuse notation $L$) with respect to the posterior density of $\bm{\beta}$ given $\bm{x},\bm{y},g^{(t)}$, we get

\begin{align}
\label{expcompll}
\mathbb{E}_{\bm{\beta}|\bm{x},\bm{y},g^{(t)}} L(g| \bm{x},\bm{y},\bm{\beta})
= \sum_{i=1}^n  \frac{\int_{\mathbb{R}^d}  \log \left[\phi_\sigma\left( y_i - x_i^\top \beta\right) g(\beta) \right]\phi_\sigma\left( y_i - x_i^\top \beta \right) g^{(t)}(\beta) d\beta}{\int_{\mathbb{R}^d}  \phi_\sigma\left( y_i - x_i^\top \beta \right) g^{(t)}(\beta)d\beta  }.
\end{align}
This is the E-step in an EM algorithm. Let $\mathcal{G}_{\text{den}}$ be the space of distributions on $\mathbb{R}^d$ for which their density functions exist. We would like to find a maximizer of \eqref{expcompll} in $\mathcal{G}_{\text{den}}$. Note that $\int_{\mathbb{R}^d} g(u) du = 1$ for $G\in\mathcal{G}_{\text{den}}$ such that $g$ is the density function of $G$, which is a restriction in the optimization. Define the Lagrangian function for the optimization as
\begin{align}
F(g;g^{(t)}) : =\mathbb{E}_{\bm{\beta}|\bm{x},\bm{y},g^{(t)}} L(g| \bm{x},\bm{y},\bm{\beta}) - n \left[ \int_{\mathbb{R}^d} g(u) du-1 \right],
\end{align}
\noindent where $n$ is the Lagrange multiplier. 
Notice that 
\begin{align}
\label{Mstep}
\argmax_{G\in\mathcal{G}_{\text{den}}} F(g;g^{(t)}) = \argmax_{G\in\mathcal{G}_{\text{den}}} \mathbb{E}_{\bm{\beta}|\bm{x},\bm{y},g^{(t)}} L(g| \bm{x},\bm{y},\bm{\beta}). 
\end{align}
To find the solution to $\eqref{Mstep}$, we can first take the derivative of $F(g;g^{(t)})$ with respect to $g$, which is a functional derivative. In general, the functional derivative of a functional $D(f)$ is defined as $\delta D/\delta f$ such that 
\begin{align}
    \lim_{\epsilon\to 0} \left[\dfrac{D[f+\epsilon\eta]-D[f]}{\epsilon}\right]= \int \dfrac{\delta D}{\delta f}(x)\eta(x)dx,
\end{align}
where $\eta$ is an arbitrary function and $\epsilon$ is a scalar value \citep{Pang1994density}. 

Taking the functional derivative of $F(g;g^{(t)})$ with respect to $g$, we obtain
\begin{align}
\frac{\delta F(g;g^{(t)})}{\delta g} = \sum_{i=1}^n  \frac{\phi_\sigma\left(y_i - x_i^\top \beta\right) g^{(t)}(\beta) }{g(\beta)\int_{\mathbb{R}^d}  \phi_\sigma\left( y_i - x_i^\top \beta \right) g^{(t)}(\beta)d\beta  } - n.
\end{align}
By letting $\frac{\delta F(g;g^{(t)})}{\delta g}=0$ and setting the solution as $g^{(t+1)}$, we recover the algorithm in \eqref{gtplus1old}:
\begin{align}
g^{(t+1)}(\beta) = \frac{1}{n}\sum_{i=1}^n  \frac{\phi_\sigma\left( y_i - x_i^\top \beta\right) g^{(t)}(\beta)}{\int_{\mathbb{R}^d}  \phi_\sigma\left( y_i - x_i^\top \beta \right) g^{(t)}(\beta)d\beta  }.
\end{align}
The following proposition states that \eqref{gtplus1old} provides the solution to the M-step.
\begin{proposition}
\label{proposition_vanilla_alg}
In the above setting,
\begin{align}
\label{mstepopt}
g^{(t+1)} = \argmax_{G\in\mathcal{G}_{\text{den}}} \mathbb{E}_{\bm{\beta}|\bm{x},\bm{y},g^{(t)}} L(g| \bm{x},\bm{y},\bm{\beta}),\;\; t=0,1,2,\cdots,
\end{align}
that is, $g^{(t+1)}$ is the unique maximizer of the expectation of the complete log-likelihood given $g^{(t)}$.
\end{proposition}

The output of this algorithm converges to NPMLE under mild assumptions, and the convergence limit $g^{(\infty)}$ can then be used as a final estimate for $g^*$. The following theorem is similar to Proposition 2 in \citet{chae2018convergence} and its proof is omitted. Recall that $\mathcal{X}$ is the support of $G^*$.
\begin{theorem}
    Suppose for every $i=1,\cdots,n$, that $\beta \rightarrow \phi_\sigma\left( y_i - x_i^\top \beta\right)$ is a continuous and strictly positive map on $\mathcal{X}$. In addition, assume that for each $\epsilon > 0 $ there exists a compact $\mathcal{X}_0 \subset \mathcal{X}$ where $\displaystyle \sup_{\beta \in \mathcal{X}_0^{\complement}} \phi_\sigma\left( y_i - x_i^\top \beta\right) < \epsilon$ for all $i=1,\cdots,n$. If a unique NPMLE $\hat{G}$ exists, then as $t \rightarrow \infty$, $G^{(t)}$ converges weakly to $\hat{G}$, where $G^{(t)}$ is the probability distribution function of $g^{(t)}$.
\end{theorem}

\subsection{EM-NPKMLE algorithm: when $G^*$ possibly does not have a density} 
\label{subsec:EM algorithm G discrete}

\begin{figure}[H]
   \centering
\includegraphics[width=0.7\textwidth]{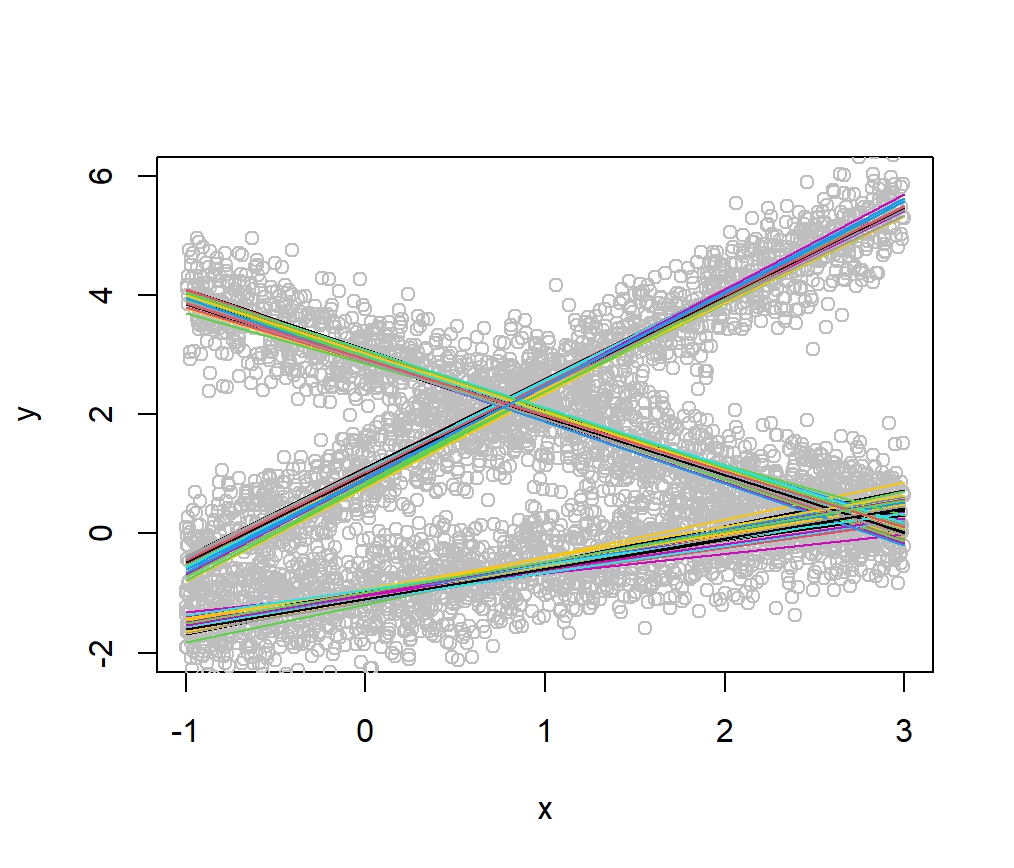} 
\caption{The regression lines of different colors representing 100 $\bm{\beta}$ coefficients sampled from an estimated density generated from the output of the EM-NPMLE algorithm for the same simulation as used in Figure \ref{fig:all_methods_intro}.}
\label{Fig:Vanilla_lots_of_lines}
\end{figure}

The success of the EM-NPMLE method in Section~\ref{subsecsec:EM alg G density} requires $G^*$ to have a density, but may not perform well when this is not true. For example, it is not expected to give a consistent estimate of the number of components when $G^*$ has a finite number of support points, because $g^{(t+1)}$ from \eqref{gtplus1old} will always be a density, although it converges to the NPMLE, which has at most $n$ support points (see Theorem 1 in~\citet{jiang2025nonparametric}). Figure \ref{Fig:Vanilla_lots_of_lines} illustrates an example of the estimated regression lines with coefficients sampled from the output of the EM-NPMLE algorithm in Section~\ref{subsecsec:EM alg G density}. The algorithm developed in \citet{jiang2025nonparametric} aiming to approximate the NPMLE also tends to overestimate the number of components in the mixture of linear regression model.

To address this issue when $G^*$ does not have a density function, for example, when it is only supported on $K$ points in $\mathbb{R}^d$, where we emphasize that we do not assume that $K$ is known, we propose to restrict the maximization of the likelihood function in \eqref{NPMLE} to a subset $\mathcal{G}_{\text{kde}}$, which is the set of the kernel density estimates (KDEs) with a bandwidth $h>0$ and a kernel function $V$ based on $n$ points. The optimization problem becomes NPKMLE in \eqref{newargmax}. The kernel function $V$ is a probability density function on $\mathbb{R}^d$ and we specifically require that it is differentiable and spherically symmetric such that $V(x)=v(\|x\|^2)$ for some $v$, where $v: \mathbb{R}_{\ge0} \to \mathbb{R}_{\ge0}$ is called the profile of $V$ \citep{comaniciu2002mean}. An example of $V$ is the density function of the standard normal distribution on $\mathbb{R}^d$.  We can write
\begin{align}
\mathcal{G}_{\text{kde}} \equiv \mathcal{G}_{\text{kde}}(v,h,n) = \left\{\frac{1}{nh^d} \sum_{\ell=1}^n v\Big(\frac{\|\cdot - \tilde\beta_\ell\|^2}{h^2}\Big):  \tilde\beta_1,\dots,\tilde\beta_n\in\mathbb{R}^d\right\}.
\end{align}
Here we fix $v$ and $h$ so that each element in $\mathcal{G}_{\text{kde}}$ is determined by $\tilde{\bm{\beta}}=\{\tilde\beta_1,\dots,\tilde\beta_n\}\subset\mathbb{R}^d$, and the corresponding distribution is denoted $G_{\bm{\beta}}$. The optimization in \eqref{newargmax} is reduced to finding the $\tilde{\bm{\beta}}$ associated with the NPKMLE, with the solution denoted by $\hat{\bm{\beta}}$. The empirical distribution based on $\hat{\bm{\beta}}$ (which is discrete) is used to estimate $G^*$ and cluster all $(x_i,y_i)$'s using a plug-in approach as described in Section~\ref{sec:intro}.

The above approach works even when the true distribution is discrete, because we can always get a kernel density estimate using a sample, even from a discrete distribution. 

To solve NPKMLE, we will again apply the EM idea. Recall that given a current estimate $G^{(t)}$ of $G^*$ with density $g^{(t)}$, the expectation of ``complete-data'' log-likelihood in \eqref{complete-log} with respect to the posterior distribution is given in \eqref{expcompll}. Since we require $g^{(t)}$ to be a KDE in $\mathcal{G}_{\text{kde}}$, $G^{(t)}$ is characterized by a set of $n$ points in $\mathbb{R}^d$ denoted by $\bm{\beta}^{(t)}=\{\beta_\ell^{(t)}\in\mathbb{R}^d: \ell=1,\cdots,n\}$, such that we can write $G^{(t)}=G_{\bm{\beta}^{(t)}}$. This is the E-step of our algorithm. For the M-step, we seek a $g^{(t+1)}$ such that

\begin{align}
\label{gtmaxofcondexp}
    g^{(t+1)} \in \argmax_{g \in \mathcal{G}_{\text{kde}}} Q(G;G^{(t)}),
\end{align}
where  
\begin{align}
Q(G;G^{(t)})=\mathbb{E}_{\bm{\beta}|\bm{x,y},G^{(t)}} L(G|\bm{ x,y,\beta}).
\end{align}
In other words, we would like to maximize the function in \eqref{expcompll} with respect to $g$ but require $g$ to take the form of 

\begin{align}
\label{gt}
    g(\beta) = g^{(t+1)}(\beta) = \frac{1}{nh^d} \sum_{\ell=1}^n v\Big(\frac{\|\beta - \beta_\ell^{(t+1)}\|^2}{h^2}\Big), 
\end{align}
where $\{\beta_\ell^{(t+1)}\in\mathbb{R}^d: \ell=1,\cdots,n\}=:\bm{\beta}^{(t+1)}$
 is a new set of points to be determined, which are viewed as an update from $\bm{\beta}^{(t)}$. In what follows we also use $\bm{\beta}^{(t)}$ as a vector by stacking all its elements together. 
 
Since any $g\in \mathcal{G}_{\text{kde}}$ is only determined by $n$ (unspecified) points in $\mathbb{R}^d$, the optimization problem in \eqref{gtmaxofcondexp} is reduced to finding these $n$ points. There is no closed form solution to this optimization problem in general, and therefore a gradient-ascent type algorithm is proposed, which features adaptive step sizes that do not need to be chosen. The gradient-ascent algorithm has no guarantee to converge to a global maximum, and hence the optimization in \eqref{gtmaxofcondexp} is understood to find a local maximum. This leads to an EM algorithm involving two loops, as given in  Algorithm \ref{Alg1} below, which we call EM-NPKMLE. 

\begin{algorithm}[ht]
\caption{EM-NPKMLE algorithm}
\label{Alg1}\vskip5pt
\textbf{Input}: $\bm{x,y}$, \; $\bm{\beta}^{(0)}=\{\beta_1^{(0)},\cdots,\beta_n^{(0)}\}$\vskip2pt
\textbf{For} $t=0,1,2,\dots$\vskip2pt
 \vskip2pt
\quad$\bm{\nu}^{(0)}=\bm{\beta}^{(t)}$\vskip2pt
\quad\quad\textbf{For} $r=0,1,2,\dots$\vskip2pt
\quad\quad\quad$\bm{\nu^{(r+1)}} \longleftarrow \xi(\bm{\nu}^{(r)};\bm{\beta^{(t)}} , \bm{x,y})$, where $\xi$ is given in \eqref{gradvec}\vskip2pt

 \vskip2pt
\quad\quad\textbf{End}\vskip2pt
\quad$\bm{\beta}^{(t+1)}\longleftarrow  \bm{\nu^{(\infty)}}$\vskip2pt
\textbf{End}\vskip2pt
\textbf{Output}: empirical distribution of $\{\beta_j^{(\infty)}: j=1,\cdots,n\}$
\end{algorithm}

The key step in the algorithm is the update from $\bm{\nu}^{(r)}$ to $\bm{\nu^{(r+1)}}$, which can be indeed understood as a gradient ascent method because it can be expressed in the following form:

\begin{align}
\label{gradientascent}
    \bm{\nu^{(r+1)}}=\xi(\bm{\nu}^{(r)};\bm{\beta^{(t)}} , \bm{x,y}):=\bm{\nu^{(r)}} + \frac{1}{C(\bm{\nu^{(r)}},\bm{\beta^{(t)}},\bm{x,y})} \nabla Q(G_{\bm{\nu^{(r)}}};G^{(t)}),
\end{align}

\noindent where $\bm{\nu}^{(r)}$ is the current update approaching to $\bm{\beta}^{(t+1)}$, $\nabla Q$ is the gradient of $Q(G_{\bm{\nu}};G^{(t)})$ with respect to $\bm{\nu}$, and $1/C(\bm{\nu^{(r)}},\bm{\beta^{(t)}},\bm{x,y})$ is an adaptive step size---see \eqref{Ceq} below for the definition of $C$.

We now derive the result in \eqref{gradientascent}. Denote
\begin{align*}
v_h(\|x\|^2):=v(\|x\|^2/h^2) \: \text{and }v_h'(\|x\|^2):=v'(\|x\|^2/h^2).
\end{align*}
Also, let $w=-v^\prime$ and $w_h=-v_h^\prime$. Plugging \eqref{gt} into \eqref{expcompll}, we get
\begin{align}
\label{Q}
&Q(G_{\bm{\nu}};G^{(t)}) \nonumber \\
= &\mathbb{E}_{\bm{\beta}|\bm{x,y},G^{(t)}} L(G_{\bm{\nu}}|\bm{ x,y,\beta}) \nonumber\\
=& \sum_{i=1}^n  \frac{\int_{\mathbb{R}^d}  \log \left[\phi_\sigma\left( y_i - x_i^\top \beta\right) \frac{1}{nh^d} \sum_{\ell=1}^n v_h(\|\beta - \nu_\ell\|^2) \right]S_i(\beta,\bm{\beta}^{(t)},\bm{x},\bm{y}) d\beta}{\int_{\mathbb{R}^d} S_i(\beta,\bm{\beta}^{(t)},\bm{x},\bm{y}) d\beta},
\end{align}
where
\begin{align}
&S_i(\beta,\bm{\beta}^{(t)},\bm{x},\bm{y})= \phi_\sigma\left( y_i - x_i^\top \beta\right)\sum_{\ell=1}^n v_h(\|\beta - \beta_\ell^{(t)}\|^2).
\end{align}
Taking the derivative of $Q$ with respect to $\nu_\ell$, $\ell=1,\cdots,n$, where $\nu_\ell$ is the $\ell$th entry of $\bm{\nu}$, we have
\begin{align}
\label{zetaeq}
\zeta(\nu_\ell;\bm{\beta^{(t)}} , \bm{x,y}) 
:= \frac{\partial Q(G_{\bm{\nu}};G^{(t)})}{\partial \nu_\ell} = \sum_{i=1}^n  \frac{\int_{\mathbb{R}^d} \frac{w_h\left(\|\beta - \nu_\ell\|^2\right)( \beta - \nu_\ell )}{\sum_{m=1}^n v_h\left(\|\beta - \nu_m\|^2\right)} S_i(\beta,\bm{\beta}^{(t)},\bm{x},\bm{y}) d\beta}{\int_{\mathbb{R}^d}  S_i(\beta,\bm{\beta}^{(t)},\bm{x},\bm{y}) d\beta  }, \;\; \ell=1,\cdots,n.
\end{align}
Setting $\zeta(\nu_\ell;\bm{\beta^{(t)}} , \bm{x,y}) =0$ we can write 
\begin{align}
\label{Aeq}
&\underbrace{\sum_{i=1}^n  \frac{\int_{\mathbb{R}^d} \frac{w_h\left(\|\beta - \nu_\ell\|^2\right)\beta}{\sum_{m=1}^n v_h\left(\|\beta - \nu_m\|^2\right)} S_i(\beta,\bm{\beta}^{(t)},\bm{x},\bm{y}) d\beta }{\int_{\mathbb{R}^d}  S_i(\beta,\bm{\beta}^{(t)},\bm{x},\bm{y}) d\beta  }}_{=A(\nu_\ell;\bm{\beta^{(t)}},\bm{x,y})} \\%
= 
\label{Ceq}
&\nu_\ell  \underbrace{\sum_{i=1}^n  \frac{\int_{\mathbb{R}^d} \frac{w_h\left(\|\beta - \nu_\ell\|^2\right)}{\sum_{m=1}^n v_h\left(\|\beta - \nu_m\|^2\right)} S_i(\beta,\bm{\beta}^{(t)},\bm{x},\bm{y}) d\beta }{\int_{\mathbb{R}^d}  S_i(\beta,\bm{\beta}^{(t)},\bm{x},\bm{y}) d\beta  }, \;\; \ell=1,\cdots,n}_{=C(\nu_\ell,\bm{\beta^{(t)}},\bm{x,y})}.
\end{align}
Or equivalently,
\begin{align}
\label{gradvec}
\xi(\nu_\ell;\bm{\beta^{(t)}} , \bm{x,y}) := \frac{A(\nu_\ell;\bm{\beta^{(t)}},\bm{x,y})}{C(\nu_\ell,\bm{\beta^{(t)}},\bm{x,y})} = 
\nu_\ell, \;\; \ell=1,\cdots,n.
\end{align}
Hence, the gradient of $Q(G_{\bm{\nu}};G^{(t)})$ with respect to $\bm{\nu}$ is 

\begin{align}
  &\nabla Q(G_{\bm{\nu}};G^{(t)}) \\
  =& \Big( \; \zeta(\nu_1;\bm{\beta^{(t)}},\bm{x,y}), \cdots, \zeta(\nu_n;\bm{\beta^{(t)}}, \bm{x,y}) \; \Big)^\top  \nonumber \\
  = & \Big( \; C(\nu_1,\bm{\beta^{(t)}},\bm{x,y}) [\xi(\nu_1;\bm{\beta^{(t)}},\bm{x,y}) - \nu_1], \cdots, C(\nu_n,\bm{\beta^{(t)}},\bm{x,y}) [\xi(\nu_n;\bm{\beta^{(t)}}, \bm{x,y}) - \nu_n]\; \Big)^\top , \nonumber 
\end{align}
which justifies \eqref{gradientascent}.
We now show that Algorithm \ref{Alg1} generates an increasing sequence of $\eqref{Q}$ as a function of $\bm{\nu^{(r)}}$. Here we require that the profile $v$ is monotonically decreasing and function $\log (\sum_{i=1}^n v(x_i))$ is convex, which is satisfied when $V$ is the Gaussian kernel since the LogSumExp function is convex.\footnote{See \url{https://www.math.uwaterloo.ca/~hwolkowi/henry/teaching/w10/367.w10/367miscfiles/pages48to60.pdf}.}

\begin{theorem}
\label{Qincreasing} 
    In the current setting, the expectation of the complete log-likelihood function $Q$ in \eqref{Q}  as a function of $\bm{\nu^{(r)}}$ is monotonically non-decreasing, i.e., for $r=0,1,\cdots$,
    \[Q(G_{\bm{\nu^{(r+1)}}};G^{(t)})\geq Q(G_{\bm{\nu^{(r)}}};G^{(t)}),\] 
and $Q(G_{\bm{\nu^{(r)}}};G^{(t)})$ converges as $r\to\infty$, where $G_{\bm{\nu^{(r+1)}}}$,  $G_{\bm{\nu^{(r)}}}$, and $G^{(t)}$ are the distributions of the kernel density estimations of $G$ using $\bm{\nu^{(r+1)}}$, $\bm{\nu^{(r)}}$, and $\bm{\beta^{(t)}}$, respectively.
\end{theorem}

The convergence of $\bm{\nu^{(r)}}$ is not implied by the above result but will be assumed in what follows. Now we show the monotonicity of the incomplete likelihood function in \eqref{incomplete-log} as a function of the sequence $G^{(t)}$.

\begin{theorem}
\label{incomp_monotonic}
    The incomplete likelihood function $L(G)$ in \eqref{incomplete-log} is bounded from above, and is a monotonically non-decreasing function of the sequence $G^{(t)}$, that is, $L(G^{(t+1)}) \geq L(G^{(t)})$, and hence $L(G^{(t)})$ converges as $t\to\infty$.
\end{theorem}

In the EM-NPKMLE algorithm, we take the sequence $\bm{\nu}_r$ in our Algorithm to convergence, which can be viewed as a complete M-step to locally maximize \eqref{gtmaxofcondexp}, and can be computationally expensive sometimes. As a lightweight variant, we may just run the gradient ascent step in the inner loop of the algorithm one time instead of taking it to convergence, which is a generalized (or gradient) EM (GEM) algorithm, and is called GEM-NPKMLE. Similar to the result in Theorem \ref{incomp_monotonic}, the incomplete likelihood function $L$ evaluated at the distribution sequence $G$ produced by GEM-NPKMLE algorithm is also non-decreasing and convergent.

\subsection{Post-processing EM-NPMLE: when $G^*$ possibly does not have a density}
\label{subsec:postprocessing}

Although the EM-NPMLE algorithm presented in Section~\ref{subsecsec:EM alg G density} to approximate NPMLE is not directly suitable for the case where the true prior distribution is discrete, if a clustering algorithm is applied as a post-processing step to a random sample generated from the density $g^{(t)}$ for $t$ sufficiently large produced in \eqref{gtplus1old}, we can still obtain a good result to estimate the components in the finite mixture of the linear regression model. The key observation is that as $t$ grows to infinity, $g^{(t)}$ gets closer to the NPMLE and the true prior distribution, formed by peaks near the true support points (see Figure \ref{Fig:Vanilla_lots_of_lines}). The clustering algorithm we choose to use in this postprocessing step is the mean shift algorithm, which does not require the specification of the number of clusters, and naturally uses local modes (peaks in density) as the central idea of clustering. The mean shift algorithm is a density-based clustering algorithm that iteratively calculates a local weighted mean starting at a sample point. A kernel function acts as a weight function for this local weighted mean, and a bandwidth determines the level of smoothing. A sequence of local weighted means is iteratively calculated until convergence, which can be shown to approximate the gradient flow of the kernel density estimator using the given sample. All sample points that converge to the same local mode are considered part of the same cluster \citep{comaniciu2003kernel}. 

The post-processing idea can be applied not only when we know the true prior distribution is discrete, but can also be extended to some other settings when we have some specific information about the distribution. For example, an interesting scenario occurs when the true prior distribution is supported on a low-dimensional submanifold in $\mathbb{R}^d$. In this case, the distribution does not have a density with respect to the Lebesgue measure, and hence does not satisfy the assumptions when we apply the EM-NPMLE algorithm in Section~\ref{subsecsec:EM alg G density}. However, we can use a random sample from the output of this EM-NPMLE algorithm as initialization and apply the subspace constrained mean shift (SCMS) algorithm \citep{ozertem2011locally,qiao2021algorithms}, which extends the mean shift algorithm to extract low-dimensional features called ridges from data---in fact, we can view a set of local modes as a 0-dimensional submanifold for which the mean shift algorithm is applicable. Like the mean shift algorithm, the SCMS algorithm also requires a kernel (weight) function and a smoothing bandwidth. However, the SCMS projects the gradient of the kernel density estimator onto a subspace spanned by a subset of the eigenvectors of the Hessian matrix using the kernel density estimator. We demonstrate this approach in a simulation study in Section~\ref{subsubsec:simulations}. 

We emphasize that the specific post-processing procedure depends on the further knowledge or assumption about the true prior distribution. \citet{jiang2025nonparametric} develop an algorithm to approximate NPMLE, and in order to improve the estimation of the clusters when the prior distribution is discrete, they add a BIC trimming step, which has a similar purpose as the mean shift post-processing step using the output of EM-NPMLE algorithm in Section ~\ref{subsecsec:EM alg G density}. 

In practice, it can be difficult to decide whether the prior distribution is continuous or discrete (see the CO2 vs GDP real data example in Section~\ref{sec:Applications}), which can create challenges of determining the best post-processing steps to use. As a comparison, the EM-NPKMLE algorithm we present in Section~\ref{subsec:EM algorithm G discrete} can be used to handle both discrete and continuous prior distributions without requiring a post-processing step. 

\section{Numerical Results}
\label{sec:Simulations and Applications}
For the EM-NPMLE algorithm introduced in Section~\ref{subsecsec:EM alg G density}, it requires an initial density $g^{(0)}$, which can be chosen as a (non-informative) uniform distribution over a bounded subset in $\mathbb{R}^d$. For the EM-NPKMLE algorithm developed in Section~\ref{subsec:EM algorithm G discrete}, we need a set of initial points $\bm{\beta}^{(0)}$, which can be chosen by sampling from the same uniform distribution. Alternatively, one can also use a random sample from the estimated density after running the EM-NPMLE algorithm. This creates initial data points that reflect the underlying distribution $G^*$ more accurately than assuming no prior knowledge of the distribution $G^*$ and using samples from an arbitrary distribution (e.g., a uniform distribution) as the starting values. Our simulation results below indicate that the choice of the initial points $\bm{\beta}^{(0)}$ does not have a strong influence on the results, as the performance of the algorithm is good for either choice described above.

We used the Gaussian kernel as the kernel function $V$ for EM-NPKMLE in all the simulations below. We propose using a bandwidth based on the maximal smoothing principle as described in \citet{terrell1990maximal}. Based on this principle, the bandwidth is chosen based on the maximum smoothing consistent with the density's estimated dispersion. We find this approach advantageous because by using such a bandwidth, we guard against choosing a smaller bandwidth that may introduce spurious features, such as additional local modes in the estimated distribution that do not exist in $G^*$. The bandwidth is defined as follows:
\begin{align}
    h_{\text{OS}}=U\times\left[\dfrac{(d+8)^{\frac{d+6}{2}}\pi^{d/2}R(V)}{16n\Gamma(\frac{d+8}{2})d(d+2)} \right]^{\frac{1}{d+4}},
\end{align}
where $R(V)=\int V^2$, which is equal to $1/(4\pi)$ when $V$ is a Gaussian kernel \citep{terrell1990maximal,davies2013scaling}. For $U$, we calculate the Gaussian-scaled interquartile range (IQR) (i.e., IQR/1.34, a robust estimator of the standard deviation of a normal density) of the values sampled from the estimated density in each dimension in our data and then take the mean, as proposed in \citet{davies2010adaptive}. We emphasize that this bandwidth selection strategy is used only when the initialization is based on sampling from the EM-NPMLE output. Otherwise an adjustment may be needed for a different initilization.

The numerical study includes simulations where the true distribution is discrete and continuous. For the simulation using a discrete distribution, when we examined the performance of the EM-NPKMLE, we tested two different initialization procedures. We also ran the GEM-NPKMLE with a single iteration in the inner loop, as discussed at the end of Section~\ref{subsec:EM algorithm G discrete}. We compared the EM-NPKMLE performance with the mean-shift algorithm as a post-processing step after sampling from the output density based on the EM-NPMLE algorithm, and with the CGM developed in \citet{jiang2025nonparametric}. In addition, we included a study verifying that cross-validation can effectively estimate $\sigma$ that was assumed to be known. For the continuous distribution case, we generated one simulation and compared the results visually using the EM--NPMKLE, the EM-NPMLE with and without a post-processing step, and the CGM. We also implemented our EM algorithms in two real applications.

We used {\sf R} version 4.1.2 \citep{R} for all the numerical results except for the CGM method in \citet{jiang2025nonparametric}, where we used the Python code referenced in their paper. 

\subsection{Simulations}
\label{subsec:simulations}

\subsubsection{Simulation 1}
\begin{figure}[ht]
    \centering
    \begin{subfigure}{0.45\textwidth}
        \centering
        \includegraphics[width=1\textwidth]{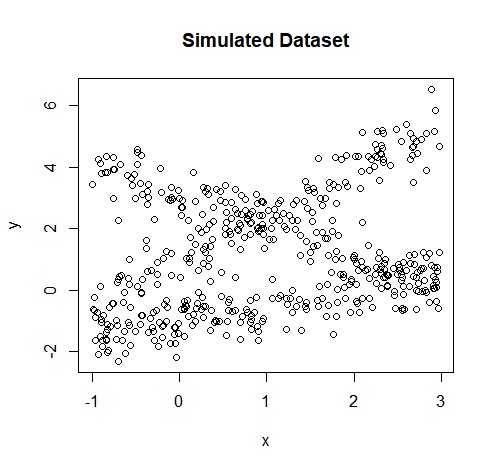} 
    \end{subfigure}\hfill
    \begin{subfigure}{0.45\textwidth}
        \centering
        \includegraphics[width=1\textwidth]{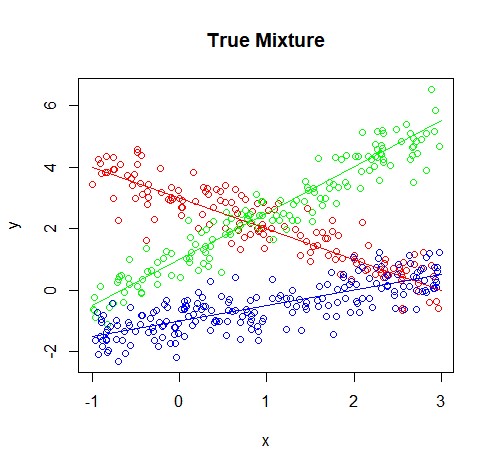} 
    \end{subfigure}
    \caption{An example of a simulated dataset from the model in \eqref{simulation_model} for $\sigma=.5$ and $n$=500 (left panel) and the corresponding true components in the mixture of linear regression model represented by different colors (right panel).}
            \label{sim1data2}
\end{figure}

We used the following model to test the algorithm, which has also been used in \citet{jiang2025nonparametric}. The mixture of linear regression model has three components with prior weights .3, .3, and .4, respectively:
\begin{align}
\label{simulation_model}
  &  y=3 - x + \varepsilon, \nonumber \\
  &  y=1 + 1.5x + \varepsilon, \\
  &  y=-1 + .5x + \varepsilon,\nonumber
\end{align}
where the noise $\varepsilon$ has a $N(0, \sigma^2$) distribution with $\sigma=.5$. The $x$ values were generated from a uniform distribution on $[-1,3]$. We ran our algorithm on 200 simulated datasets (see Figure \ref{sim1data2} for an example).  We assume that $\sigma$ is known until we present the result using cross-validation. 

In our simulation experiment, we used the uniform distribution over $[-4,4]^2$ as the initialization for the EM-NPMLE algorithm, to iteratively calculate $g^{(t+1)}$ using the expression in \eqref{gtplus1old} until the $\mathbb{L}^2$ distance between $g^{(t)}$ and $g^{(t+1)}$ fell below a predefined small threshold. The final estimated density was used to sample $n$ points as the starting points for the EM-NPKMLE algorithm. We stepped through sample sizes starting from 500 until 10,000.
The EM-NPKMLE algorithm produced a set of points that defines a KDE solution to \eqref{gtmaxofcondexp}. Suppose its empirical distribution can be represented by $\hat G = \sum_{j=1}^{\hat K} \hat\pi_j \delta_{\hat\beta_j}$ where the $\hat\beta_j$'s are all distinct, that is, we aggregate the solution points according to their unique values. For a mixture of linear regression model that has $K$ components with $K$ unknown, the true prior distribution $G^* = \sum_{j=1}^K \pi_j \delta_{\beta_j}$ is estimated by $\hat G$, where each $\beta_j$ is estimated by the closest $\hat\beta_j$, and each $\pi_j$ by the associated $\hat \pi_j$. To cluster a pair of points $(x_i,y_i)$, it is assigned to a class $j$ for which the posterior probability is maximized, i.e., 
\begin{align}
\label{bayesposterior}
\argmax_{j} \dfrac{\hat{\pi}_j\phi_\sigma\left( y_i - x_i^\top \hat{\beta}_j\right)}{\sum_{k=1}^{\hat K}\hat{\pi}_k\phi_\sigma\left( y_i - x_i^\top \hat{\beta}_k \right)}.
\end{align}
We report the average adjusted Rand index \citep{hubert1985comparing}, the proportion of samples that were correctly identified as having three components, and the average Wasserstein-2 distance~\citep{nguyen2011wasserstein} between the probability measures $\hat{G}$ produced by the algorithm in question and $G^*$, i.e., 
\begin{align}
\label{Wasserstein}
W_2(G^*,\hat{G}):= \left[\min\limits_{\bm{\gamma} \in \Gamma(\bm{\pi},\bm{\hat{\pi}})} \sum^K_{i=1} \sum^{\hat K}_{j=1}||\beta_i-\hat{\beta}_j||^{2}\gamma_{ij} \right]^{1/2},
\end{align}
where $\bm{\gamma}=(\gamma_{ij})_{1\le i\leq K, 1\le j\leq \hat K} \in [0,1]^{K\times \hat K}$,  $\Gamma(\bm{\pi},\bm{\hat{\pi}})$ is the set of all matrices $\bm{\gamma}$ satisfying $\sum^K_{i=1} \gamma_{ij}=\hat{\pi}_j$, $\sum^{\hat K}_{j=1} \gamma_{ij}=\pi_i$, where $\bm{\pi}=\{\pi_1,\cdots,\pi_K\}$ are the true weights, and $\bm{\hat{\pi}}=\{\hat{\pi}_1,\cdots,\hat{\pi}_{\hat K}\}$ are the estimated weights; and $\beta_i$'s are the true coefficients and $\hat{\beta_j}$'s are the estimated coefficients (see Table \ref{sim2sum}). 

Additionally, note that some points at the intersections of the linear regression components will unavoidably be mislabeled using \eqref{bayesposterior} by any estimators of $G^*$, which results in an adjusted Rand index less than 1. As an oracle reference for our algorithm, we calculated the average adjusted Rand index when $G$ was specified as $G^*$, and the posterior probabilities were directly calculated. We also report the bias and standard deviation of the estimated coefficients and weights when three components, the true number of components, were estimated. 

\paragraph{EM-NPKMLE:} For the results of the EM-NPKMLE algorithm, see Table \ref{sim2sum} for information on the proportion of simulations in which the true number of components were detected, and Table \ref{sim2sum2} for bias and standard deviation of the errors. Standard deviations are also reported for all values (in parentheses), where applicable. 

\renewcommand{\arraystretch}{0.6}
\begin{table}[ht!]
\centering
\caption{Results of Simulation 1 using EM-NPKMLE algorithm.}
\label{sim2sum}
\resizebox{\columnwidth}{!}{\begin{tabular}{cccccc}
\hline
& Sample Size & Avg Adj RI & Avg Adj RI G* & Prop of Sim with 3 Comp & Avg W-2 Distance\\ 
 \hline
& 500  & 0.415 (0.119)& 0.665 (0.033)  & 0.540 & 0.818 (0.137)   \\ 
& 1000 & 0.567 (0.034)& 0.663 (0.023) & 0.990 & 0.562 (0.097)  \\ 
& 5000 & 0.645 (0.011)& 0.664 (0.010) & 0.995 & 0.343 (0.080)  \\ 
& 10000 & 0.651 (0.008)& 0.663 (0.008) & 0.995  & 0.288 (0.081)  \\ 
   \hline
\end{tabular}}
\vspace*{2mm}
\end{table}

\renewcommand{\arraystretch}{0.7}
\begin{table}[ht!]
\centering
\caption{Results of Simulation 1 using EM-NPKMLE algorithm.}
\label{sim2sum2}
\resizebox{\columnwidth}{!}{\begin{tabular}{ccccc}
\hline
Sample Size & Estimate & 1st Comp & 2nd Comp & 3rd Comp\\ 
 \hline
500 
& $\beta$: bias $\pm$ SD  & (-0.360, 0.227)$\pm$(0.053, 0.074)  & (-0.565, -0.363)$\pm$(0.234, 0.138)   & (0.460, 0.230)$\pm$(0.119, 0.077)  \\ 
& $\pi$: bias $\pm$ SD  & -0.005$\pm$0.024 & -0.026$\pm$0.059 & 0.031$\pm$0.056 \\ 
& & & & \\

1000 
& $\beta$: bias $\pm$ SD & (-0.291, 0.157)$\pm$(0.037, 0.042) & (-0.397, -0.164)$\pm$(0.112, 0.062) & (0.297, 0.143)$\pm$(0.048, 0.046) \\ 
& $\pi$: bias $\pm$ SD & 0.002$\pm$0.019 & -0.003$\pm$0.017 & 0.001$\pm$0.019 \\ 
& & & & \\

5000 
& $\beta$: bias $\pm$ SD & (-0.155, 0.033)$\pm$(0.014, 0.016) & (-0.157, 0.003)$\pm$(0.032, 0.018) & (0.149, 0.038)$\pm$(0.019, 0.014) \\ 
& $\pi$: bias $\pm$ SD & -0.001$\pm$0.008 & 0$\pm$0.007 & 0.001$\pm$0.009 \\ 
& & & & \\

10000 
& $\beta$: bias $\pm$ SD & (-0.112, 0.004)$\pm$(0.011, 0.010) & (-0.115, 0.013)$\pm$(0.018, 0.012) & (0.113, 0.027)$\pm$(0.013, 0.010) \\ 
& $\pi$: bias $\pm$ SD & -0.001$\pm$0.006 & 0$\pm$0.006 & 0.001$\pm$0.006 \\ 
   \hline
\end{tabular}}
\vspace*{2mm}
\end{table}

The results suggest that our algorithm produces consistent estimates of the $\beta$ components and weights, as well as a consistent estimator of the number of components. As $n$ increases, the values of these estimates approach the true parameter values and true number of components. As $n$ gets larger, we observe that all of the points that are mislabeled are at the intersections of the components, and the slightly worse performance of the estimated $G$ from our algorithm than using $G^*$ is due to the bias of the $\beta$ estimates creating intersection areas that are slightly larger than when $G$ is set to $G^*$. Since we use the bandwidth based on the oversmoothing principle, we accept slightly larger bias in exchange for avoiding detecting spurious modes (components) in the data.

\FloatBarrier

In what follows we compare the above results using the EM-NPKMLE algorithm with those using the the EM-NPMLE algorithm with mean shift as a post-processing step as described in Section \ref{subsec:postprocessing} and the conditional gradient method (CGM) as described in \citet{jiang2025nonparametric}. We stepped through the same sequence of sample sizes as in Tables \ref{sim2sum} and \ref{sim2sum2}.

\paragraph{EM-NPMLE and mean-shift post-processing:} As shown in Tables \ref{sim2summs} and \ref{sim2summs2}, the mean shift algorithm as a post-processing step after EM-NPMLE produced results that suggest consistent estimation of the true $\beta$ values and weights, as well as the true number of components. When using the mean shift for post-processing, we also used the bandwidth based on the maximal smoothing principle as described previously. 

\paragraph{CGM algorithm in \citet{jiang2025nonparametric}:} As shown in Tables \ref{sim2summsjiang} and \ref{sim2summsjiang2}, the results of the CGM algorithm do not suggest that this algorithm can consistently estimate the number of modes, though the decreasing average W-2 distance with increasing sample size supports the idea of convergence in distribution between the true and estimated distributions.

\renewcommand{\arraystretch}{0.6}
\begin{table}[ht!]
\centering
\caption{Results of Simulation 1 using EM-NPMLE and mean-shift post-processing.}
\label{sim2summs}
\resizebox{\columnwidth}{!}{\begin{tabular}{cccccc}
\hline
& Sample Size & Avg Adj RI & Avg Adj RI G* & Prop of Sim with 3 Comp & Avg W-2 Distance\\ 
 \hline
& 500  & 0.620 (0.094) & 0.665 (0.033)  & 0.915  & 0.612 (0.238)    \\ 
& 1000 & 0.657 (0.024)& 0.663 (0.023)  & 0.995 & 0.465 (0.141)   \\ 
& 5000 & 0.663 (0.010)& 0.664 (0.010)  & 0.990 & 0.311 (0.092)  \\ 
& 10000 & 0.662 (0.008)& 0.663 (0.008) & 1  & 0.265 (0.089)  \\ 
   \hline
\end{tabular}}
\vspace*{2mm}
\end{table}

\renewcommand{\arraystretch}{0.7}
\begin{table}[ht!]
\centering
\caption{Results of Simulation 1 using EM-NPMLE and mean-shift post-processing.}
\label{sim2summs2}
\resizebox{\columnwidth}{!}{\begin{tabular}{ccccc}
\hline
Sample Size & Estimate & 1st Comp & 2nd Comp & 3rd Comp\\ 
 \hline
500 
& $\beta$: bias $\pm$ SD  & (-0.012, 0.034)$\pm$(0.062, 0.040)  & (-0.173, -0.080)$\pm$(0.108, 0.045)   & (0.088, 0.031)$\pm$(0.062, 0.042) \\ 
& $\pi$: bias $\pm$ SD  & -0.001$\pm$0.025 & -0.002$\pm$0.027 & 0.030$\pm$0.028 \\ 
& & & & \\

1000 
& $\beta$: bias $\pm$ SD & (-0.002, 0)$\pm$(0.049, 0.031) & (-0.077, -0.021)$\pm$(0.061, 0.029) & (0.041, 0.013)$\pm$(0.044, 0.026) \\ 
& $\pi$: bias $\pm$ SD & 0.002$\pm$0.019 & -0.002$\pm$0.018 & 0$\pm$0.019 \\ 
& & & & \\

5000 
& $\beta$: bias $\pm$ SD & (0, 0)$\pm$(0.021, 0.015) & (-0.009, 0.005)$\pm$(0.023, 0.012) & (0.013, -0.003)$\pm$(0.018, 0.013) \\ 
& $\pi$: bias $\pm$ SD & -0.001$\pm$0.008 & 0$\pm$0.008 & 0.001$\pm$0.009 \\ 
& & & & \\

10000 
& $\beta$: bias $\pm$ SD & (0, 0)$\pm$(0.014, 0.010) & (-0.007, 0.004)$\pm$(0.017, 0.011) & (0.010, -0.002)$\pm$(0.012, 0.008) \\ 
& $\pi$: bias $\pm$ SD & -0.001$\pm$0.006 & 0$\pm$0.006 & 0$\pm$0.006 \\ 

   \hline
\end{tabular}}
\vspace*{2mm}
\end{table}

\renewcommand{\arraystretch}{0.6}
\begin{table}[ht!]
\centering
\caption{Results of Simulation 1 using CGM.}
\label{sim2summsjiang}
\begin{tabular}{ccccc}
\hline
& Sample Size & Avg Adj RI & Avg Adj RI G* & Prop of Sim with 3 Comp \\ 
 \hline
& 500  & 0.590 (0.066)  & 0.665 (0.033)  & 0      \\ 
& 1000 & 0.581 (0.068) & 0.663 (0.023)  & 0      \\ 
& 5000 & 0.602 (0.053)  & 0.664 (0.010)  & 0    \\ 
& 10000 & 0.596 (0.057) & 0.663 (0.008) &  0  \\ 
   \hline
\end{tabular}
\vspace*{2mm}
\end{table}

\renewcommand{\arraystretch}{0.6}
\begin{table}[ht!]
\centering
\caption{Results of Simulation 1 using CGM.}
\label{sim2summsjiang2}
\begin{tabular}{cccc}
\hline
& Sample Size & Avg W-2 Distance & Mean No. of Comp\\ 
 \hline
& 500  & 0.439 (0.109)  & 9.295 (2.095)    \\ 
& 1000 & 0.373 (0.093)  & 8.550 (1.744) \\ 
& 5000 & 0.260 (0.069) & 7.800 (1.504)  \\ 
& 10000 & 0.229 (0.064) & 7.570 (1.274)  \\ 
   \hline
\end{tabular}
\vspace*{2mm}
\end{table}

\paragraph{EM-NPKMLE with uniform initialization:} We also ran our EM-NPKMLE algorithm on the same model as in \eqref{simulation_model} using initial points drawn from a uniform distribution over $[-4,4]^2$ to demonstrate that our algorithm does not depend on initial points sampled from the estimated density $g$ derived from the EM-NPKMLE algorithm. We currently do not have a bandwidth selection strategy for the uniform initialization, and we manually tuned the bandwidth by using the maximal smoothing principle multiplied by a constant $c=1.15$. As shown in Tables  \ref{sim2sumuniform1} and \ref{sim2sumuniform2}, the results are comparable to running our EM-NPKMLE algorithm using the output of EM-NPMLE for the initialization.

\renewcommand{\arraystretch}{0.6}
\begin{table}[ht!]
\centering
\caption{Results of Simulation 1 using EM-NPKMLE Algorithm with uniform initialization.}
\label{sim2sumuniform1}
\resizebox{\columnwidth}{!}{\begin{tabular}{cccccc}
\hline
& Sample Size & Avg Adj RI & Avg Adj RI G* & Prop of Sim with 3 Comp & Avg W-2 Dist\\ 
 \hline
& 500  & 0.146 (0.043) & 0.665 (0.033) & 0.030 &  1.130 (0.059)     \\ 
& 1000  & 0.283 (0.140) & 0.663 (0.023)   &  0.515   & 1.020 (0.114)        \\ 
& 5000 & 0.624 (0.012) & 0.664 (0.010)   &   1  &  0.422 (0.060)   \\ 
& 10000  & 0.642 (0.008) & 0.663 (0.008)   &  1 &  0.320 (0.052) \\ 
   \hline
\end{tabular}}
\vspace*{2mm}
\end{table}

\renewcommand{\arraystretch}{0.7}
\begin{table}[ht!]
\centering
\caption{Results of Simulation 1 using EM-NPKMLE Algorithm with uniform initialization.}
\label{sim2sumuniform2}
\resizebox{\columnwidth}{!}{\begin{tabular}{ccccc}
\hline
Sample Size & Estimate & 1st Comp & 2nd Comp & 3rd Comp\\ 
 \hline
500 
& $\beta$: bias $\pm$ SD  & (-0.453, 0.273)$\pm$(0.035, 0.072)   & (-0.888, -0.504)$\pm$(0.259, 0.149) & (0.643, 0.325)$\pm$(0.136, 0.070) \\
& $\pi$: bias $\pm$ SD  & 0.029$\pm$0.017 & -0.039$\pm$0.211 & 0.010$\pm$0.218  \\
& & & & \\

1000 
& $\beta$: bias $\pm$ SD  & (-0.430, 0.333)$\pm$(0.039, 0.048)  & (-0.717, -0.364)$\pm$(0.211, 0.094)    & (0.556, 0.299)$\pm$(0.094, 0.075)   \\
& $\pi$: bias $\pm$ SD  & 0.049$\pm$0.018 & -0.119$\pm$0.066 & 0.070$\pm$0.063 \\
& & & & \\

5000 
& $\beta$: bias $\pm$ SD  & (-0.224, 0.101)$\pm$(0.013, 0.018)  & (-0.234, -0.029)$\pm$(0.036, 0.020)  & (0.222, 0.073)$\pm$(0.020, 0.016)  \\ 
& $\pi$: bias $\pm$ SD  & 0.009$\pm$0.007  & -0.018$\pm$0.006   & 0.009$\pm$0.007 \\
& & & & \\ 

10000 
& $\beta$: bias $\pm$ SD  & (-0.164, 0.047)$\pm$(0.010, 0.011)  & (-0.164, 0.001)$\pm$(0.019, 0.013)   & (0.164, 0.043)$\pm$(0.014, 0.011)   \\ 
& $\pi$: bias $\pm$ SD  & 0$\pm$0.005  & -0.008$\pm$0.004  & 0.008$\pm$0.005   \\  
   \hline 
\end{tabular}}
\vspace*{2mm}
\end{table}

\paragraph{GEM-NPKMLE:}As discussed at the end of Section~\ref{subsec:EM algorithm G discrete}, GEM-NPKMLE is a variant of EM-NPKMLE, where we run a single iteration in the inner loop instead of running it to convergence. This approach saves computation time. For example, for the simulation with n=5000 and the EM-NPMLE initialization, the EM-NPKLME took an average of 115.888 (40.759) minutes, while the GEM-NPKMLE took an average of 15.448 (2.194) minutes. It also produced similar results compared to Tables \ref{sim2sum}, \ref{sim2sum2}, \ref{sim2summs}, and \ref{sim2summs2} for larger sample sizes, as can be seen in Tables \ref{simGEM2} and \ref{simGEM21}. This suggests that the GEM-NPKMLE approach is viable when saving computation time is important and there is a sufficient sample size. The initial points we used for GEM-NPKMLE were sampled from the output of EM-NPMLE, which was the same strategy that was used in the first part of our simulations. However, in this case, we multiplied the initial bandwidth by a constant $c=1.2$ to improve the algorithm's performance. 

\renewcommand{\arraystretch}{0.6}
\begin{table}[ht!]
\centering
\caption{Results of Simulation 1 using GEM-NPKMLE algorithm.}
\label{simGEM2}
\resizebox{\columnwidth}{!}{\begin{tabular}{cccccc}
\hline
& Sample Size & Avg Adj RI & Avg Adj RI G* & Prop of Sim with 3 Comp & Avg W-2 Dist\\ 
 \hline
& 500  & 0.133 (0.034) & 0.665 (0.033) & 0.235 &  1.070 (0.062)   \\ 
& 1000  & 0.213 (0.106) & 0.663 (0.023)   & 0.195  &  0.993 (0.059)  \\ 
& 5000 &  0.614 (0.012) & 0.664 (0.010)   & 0.840   & 0.401 (0.066)   \\ 
& 10000  & 0.637 (0.008) & 0.663 (0.008)   & 0.930  &  0.323 (0.070)   \\ 
   \hline
\end{tabular}}
\vspace*{2mm}
\end{table}

\renewcommand{\arraystretch}{0.7}
\begin{table}[ht!]
\centering
\caption{Results of Simulation 1 using GEM-NPKMLE algorithm.}
\label{simGEM21}
\resizebox{\columnwidth}{!}{\begin{tabular}{ccccc}
\hline
Sample Size & Estimate & 1st Comp & 2nd Comp & 3rd Comp\\ 
 \hline
500 
& $\beta$: bias $\pm$ SD  & (-0.527,0.360)$\pm$(0.067, 0.083) & (-0.983,-0.642)$\pm$(0.135,0.071)  & (0.839,0.387)$\pm$(0.071,0.067)     \\ 
& $\pi$: bias $\pm$ SD  & 0.005$\pm$(0.027)   & -0.298$\pm$(0.002) & 0.292$\pm$(0.027)  \\ 
& & & & \\

1000 
& $\beta$: bias $\pm$ SD  &  (-0.414,0.308)$\pm$(0.052,0.059)   &  (-1.052,-0.551)$\pm$(0.175,0.079)  & (0.757,0.402)$\pm$(0.092,0.075)    \\  
& $\pi$: bias $\pm$ SD  & 0.007$\pm$0.019  & -0.291$\pm$0.048   & 0.284$\pm$0.042   \\  
& & & & \\

5000 
& $\beta$: bias $\pm$ SD  & (-0.241,0.104)$\pm$(0.014,0.018)   &  (-0.282,-0.069)$\pm$(0.041,0.023)   &  (0.231,0.088)$\pm$(0.020,0.016)  \\  
& $\pi$: bias $\pm$ SD  & -0.001$\pm$0.008  & 0$\pm$0.007   & 0.001$\pm$0.009 \\ 
& & & & \\  

10000 
& $\beta$: bias $\pm$ SD  & (-0.184,0.056)$\pm$(0.010,0.011)   & (-0.196,-0.009)$\pm$(0.022,0.013)   & (0.172,0.053)$\pm$(0.014,0.011) \\  
& $\pi$: bias $\pm$ SD  & -0.001$\pm$(0.006)   &  0$\pm$(0.006)   & 0.001$\pm$(0.006)   \\  
   \hline 
\end{tabular}}
\vspace*{2mm}
\end{table}

\paragraph{Cross-validation for unknown $\sigma$:} 

If $\sigma$ is unknown, we can implement a cross-validation (CV) approach found in \citet{jiang2025nonparametric} to estimate the parameter $\sigma$. First we divide the dataset $T=\{(x_i,y_i):\; i=1,\dots,n\}$ into $C$ folds: $T_1,\dots,T_C$. For each $c\in\{1,2,...C\}$, one fold $T_c$ is left out as test data and the remaining folds are used to obtain an estimator $\hat{G}^{-c}$ for $G^*$. Let 
\begin{align*}
f_{x_i}^{\hat{G}^{-c}}(y_i) =\int \phi_\sigma\left( y_i - x_i^\top \beta\right) d\hat{G}^{-c}(\beta), \;\; i=1,\cdots,n.
\end{align*}
The $\sigma$ value that minimizes the following objective function is selected and denoted by $\hat\sigma$:
\begin{align*}
CV(\sigma)= -\sum_{c=1}^C \sum_{(x_i,y_i) \in T_c} \log(f_{x_i}^{\hat{G}^{-c}}(y_i)).
\end{align*}

In our simulation, we used a random sample from the output of the EM-NPMLE algorithm as the initialization for our GEM-NPKMLE algorithm (with the oversmoothing bandwidth) to calculate $\hat{G}^{-c}$. We tested this method on the model in \eqref{simulation_model} with $C=5$ to get $\hat\sigma$, which was further plugged into the EM-NPKMLE algorithm. The results are provided in Tables \ref{simEM1_crossval} and \ref{simEM12_crossval}, which are comparable to those in Tables \ref{sim2sum} and \ref{sim2sum2}.

\renewcommand{\arraystretch}{0.6}
\begin{table}[ht!]
\centering
\caption{Results of Simulation 1 using EM-NPKMLE algorithm and cross-validation to estimate $\sigma$.}
\label{simEM1_crossval}
\resizebox{\columnwidth}{!}{\begin{tabular}{cccccc}
\hline
& Sample Size & Avg Adj RI & Avg Adj RI G* & Prop of Sim with 3 Comp & Avg W-2 Dist\\ 
 \hline
& 500  & 0.302 (0.204) & 0.665 (0.033) & 0.390 & 0.891 (0.192)     \\ 
& 1000  & 0.574 (0.045)  & 0.663 (0.023) & 0.975 & 0.557 (0.118)      \\ 
& 5000 & 0.645 (0.011)  & 0.664 (0.010)  & 0.995 & 0.343 (0.080)   \\ 
& 10000 & 0.651 (0.008)& 0.663 (0.008) & 0.995  & 0.288 (0.081)  \\ 
   \hline
\end{tabular}}
\end{table}

\renewcommand{\arraystretch}{0.7}
\begin{table}[ht!]
\centering
\caption{Results of Simulation 1 using EM-NPKMLE algorithm and cross-validation to estimate $\sigma$.}
\label{simEM12_crossval}
\resizebox{\columnwidth}{!}{\begin{tabular}{ccccc}
\hline
Sample Size & Estimate & 1st Comp & 2nd Comp & 3rd Comp\\ 
 \hline

500
& $\beta$: bias $\pm$ SD & (-0.328,0.201)$\pm$(0.080,0.090) & (-0.425,-0.281)$\pm$(0.236,0.177)  &  (0.375,0.193)$\pm$(0.162,0.091) \\ 
& $\pi$: bias $\pm$ SD & -0.010$\pm$0.025 &  -0.028$\pm$0.071 & 0.039$\pm$0.066 \\ 
& & & & \\

1000 
& $\beta$: bias $\pm$ SD  & (-0.273, 0.143)$\pm$(0.048, 0.047) & (-0.355,-0.151)$\pm$(0.137,0.068) & (0.271,0.136)$\pm$(0.073,0.047) \\ 
& $\pi$: bias $\pm$ SD  & -0.003$\pm$0.018 &  -0.002$\pm$0.018 & 0.005$\pm$0.019  \\   
& & & & \\

5000 
& $\beta$: bias $\pm$ SD & (-0.155, 0.033)$\pm$(0.0143, 0.016)   & (-0.157,0.003)$\pm$(0.032,0.018) & (0.149,0.038)$\pm$(0.019,0.014)  \\ 
& $\pi$: bias $\pm$ SD & -0.001$\pm$0.008  & 0$\pm$0.007  & 0.001$\pm$0.009  \\ 
& & & & \\

10000 
& $\beta$: bias $\pm$ SD & (-0.112, 0.004)$\pm$(0.011, 0.010) & (-0.115, 0.013)$\pm$(0.018, 0.012) & (0.113, 0.027)$\pm$(0.013, 0.010) \\ 
& $\pi$: bias $\pm$ SD & -0.001$\pm$0.006 & 0$\pm$0.006 & 0.001$\pm$0.006 \\ 
   \hline

\end{tabular}}

\vspace*{2mm}
\end{table}

\subsubsection{Simulation 2}
\label{subsubsec:simulations}
\begin{figure}[ht!]
    \centering
    \begin{subfigure}[b]{0.48\textwidth}
        \centering
        \includegraphics[width=\textwidth]{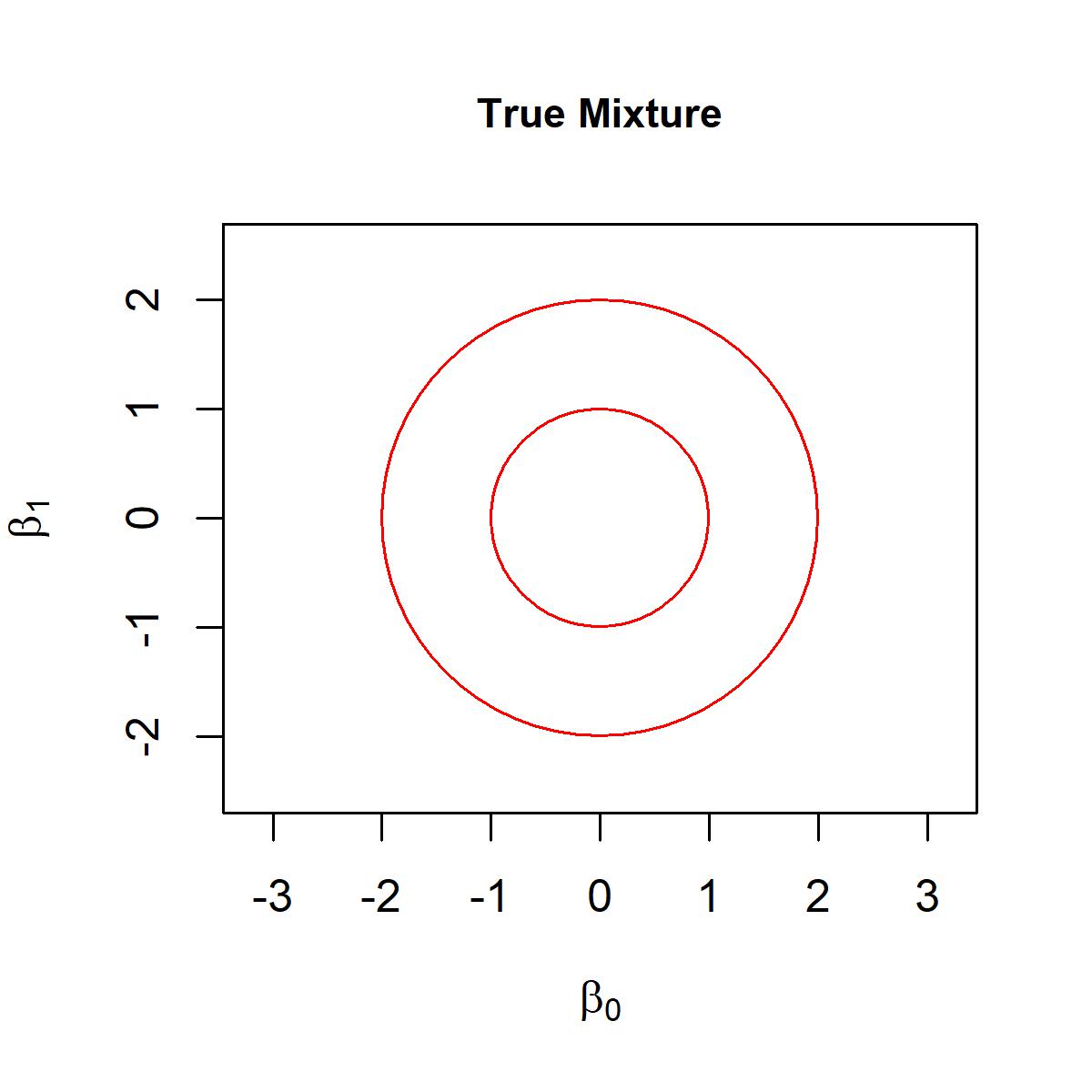}
        \label{sim2truemix}
    \end{subfigure}
    \hfill
    \begin{subfigure}[b]{0.48\textwidth}
        \centering
        \includegraphics[width=\textwidth]{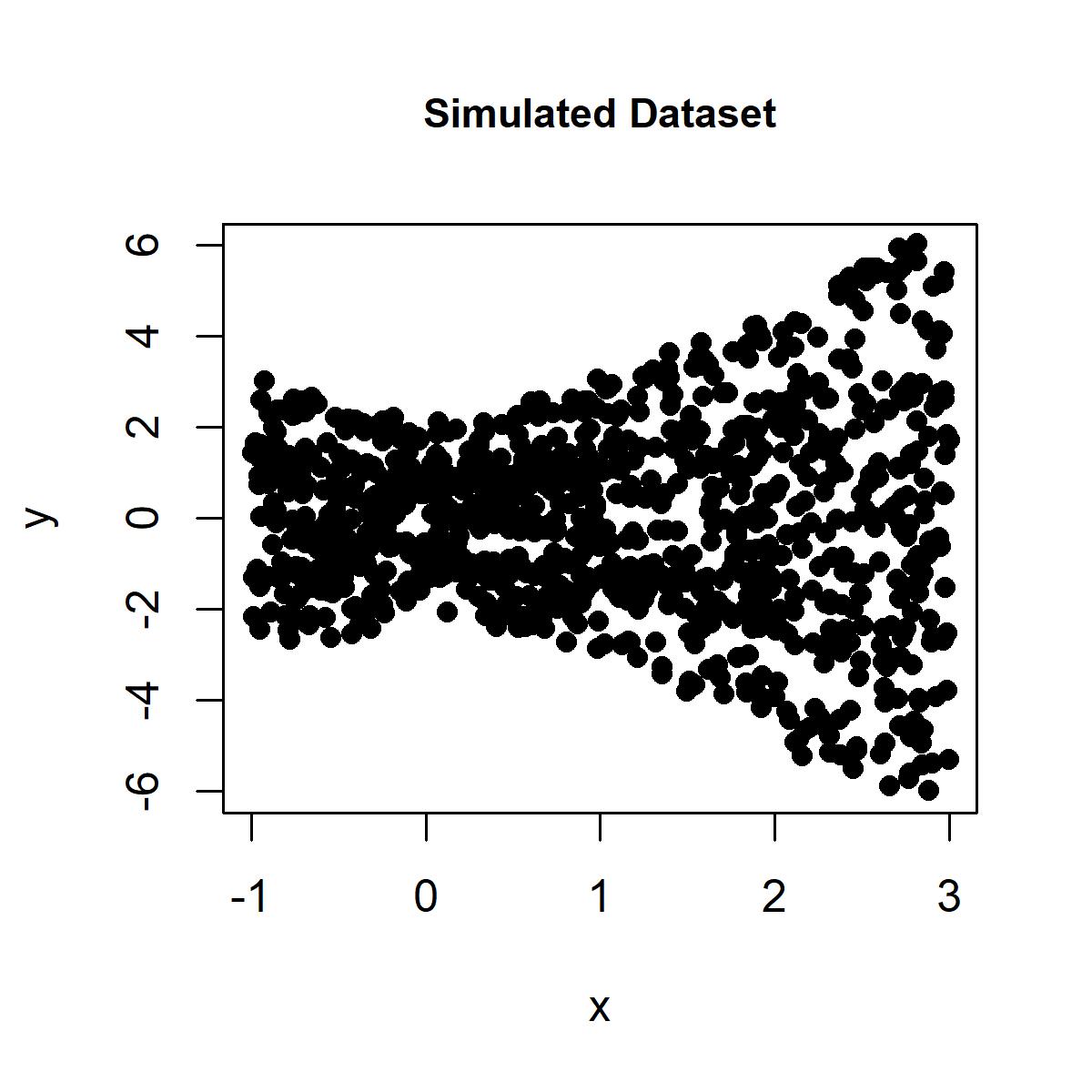}
        \label{simdataset2}
    \end{subfigure}
    \caption{Left panel: the support of $G^*$ in Simulation 2; Right panel: the sample used in Simulation 2. }
    \label{fig:sim2comparison}
\end{figure}

Unlike the EM-NPMLE algorithm, our EM-NPKMLE algorithm does not assume the knowledge of a particular structure in the true distribution, such as a low-dimensional manifold as the support. To further demonstrate this, we use a continuous distribution as the model, which consists of a bivariate distribution represented by a mixture of uniform distributions over two concentric circles, each with center at the origin (see Figure \ref{fig:sim2comparison}). The model is given in \eqref{crmodel}, where $\sigma=.2$ is assumed to be known, and the true distribution $G^*$ is 
\begin{align*}
    \frac{1}{2}\times\text{Uniform}\{\mathcal{B}(1)\} + \frac{1}{2}\times\text{Uniform}\{\mathcal{B}(2)\},
\end{align*}
where $\mathcal{B}(r)=\{\beta \in \mathbb{R}^2:||\beta||=r\}$. This model has also been used in \citep{jiang2025nonparametric}.

We uniformly sampled 10000 design points $x_i$'s from $[-1,3]$. For our EM-NPKMLE algorithm, we used the output of the EM-NPMLE algorithm as the initial $\bm{\beta}^{(0)}$ points and used the oversmoothing bandwidth. We got similar results when we used points sampled from a uniform distribution on $[-4,4]^2$ as the initial $\bm{\beta}^{(0)}$ and applied a constant to the oversmoothing bandwidth. 

In this simulation we also tested the EM-NPMLE algorithm with and without SCMS as a post-processing step, as described in 
Section~\ref{subsec:postprocessing}. The EM-NPMLE algorithm was initialized by a uniform density over $[-4,4]^2$. Then the SCMS algorithm was applied to 10000 $\beta$ points sampled from the estimated density produced by the EM-NPMLE algorithm. The bandwidth used for the SCMS algorithm was selected by the biased cross-validation method aiming to minimize an asymptotic mean integrated squared error (AMISE) estimate between the estimated and true density gradients \citep{sain1994cross}. This was implemented by the `Hbcv.diag' function in the {\sf R} package {\sf ks} \citep{ks}, and we used the mean of the diagonal entries in the selected diagonal bandwidth matrix.

\begin{figure}[!htbp]
    \centering

    \begin{subfigure}[b]{0.45\textwidth}
        \centering
        \includegraphics[width=\textwidth]{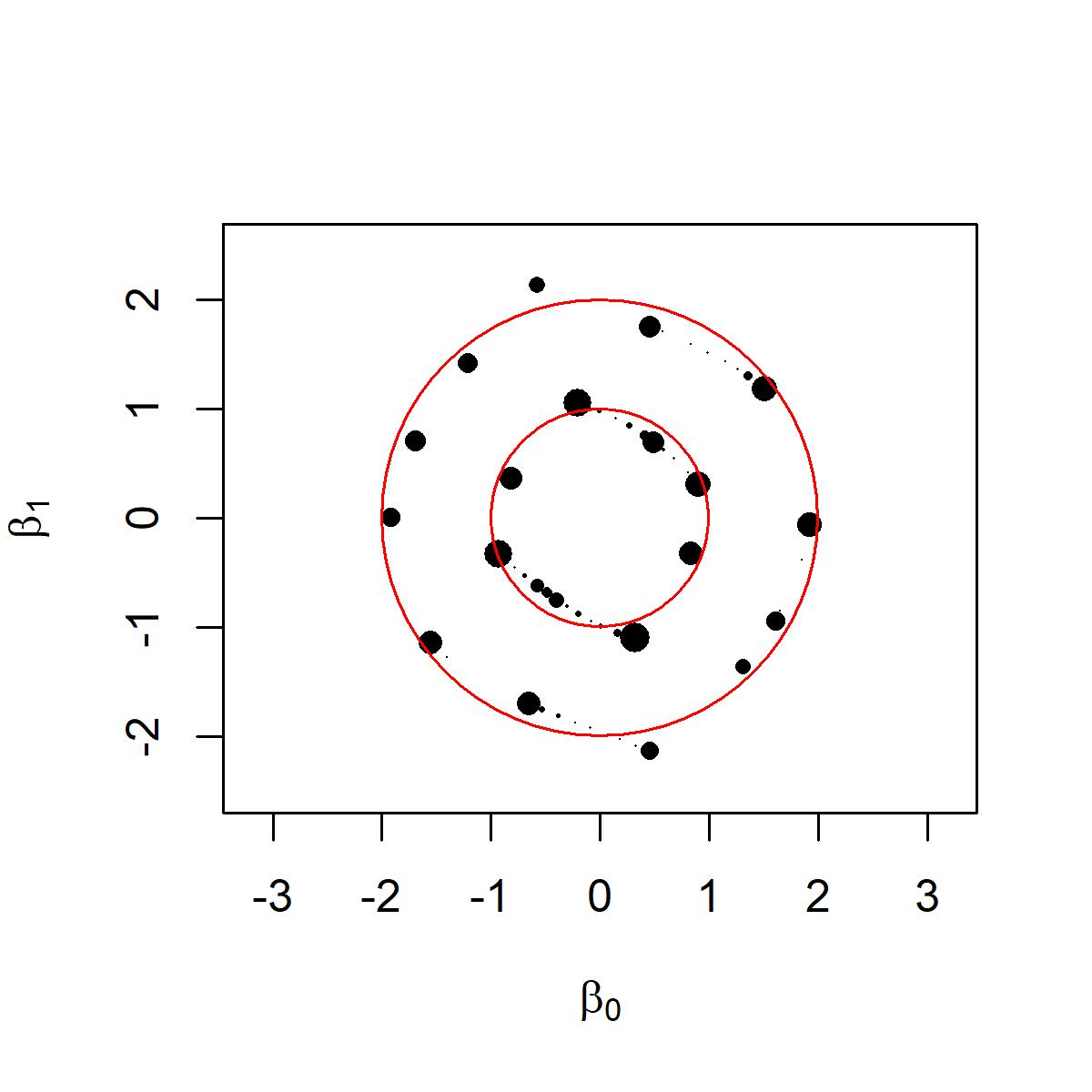}
        \caption{}
        \label{fig:em}
    \end{subfigure}
    \hfill
    \begin{subfigure}[b]{0.45\textwidth}
        \centering
        \includegraphics[width=\textwidth]{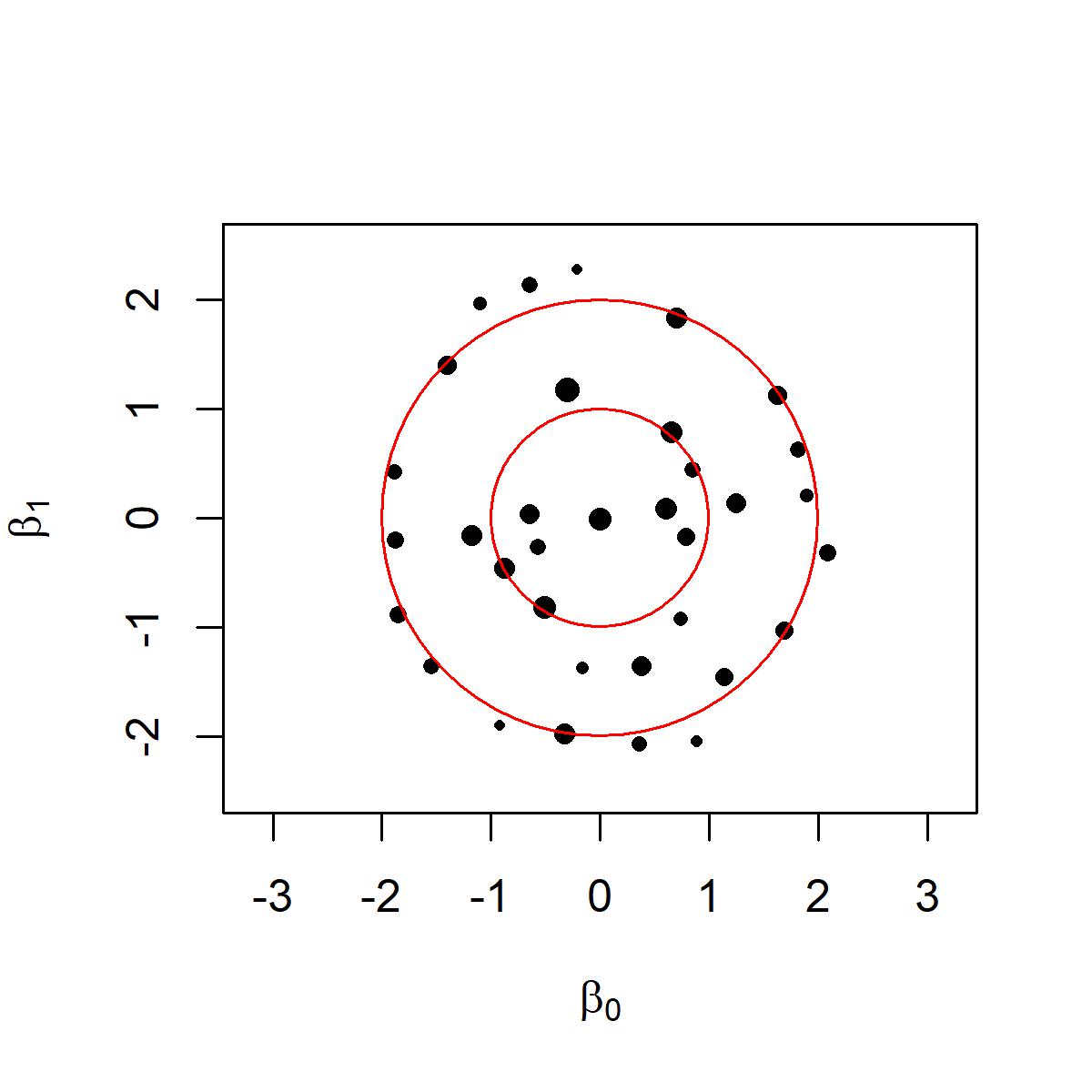}
        \caption{}
        \label{fig:cgm}
    \end{subfigure}
    
    \begin{subfigure}[b]{0.45\textwidth}
        \centering
        \includegraphics[width=\textwidth]{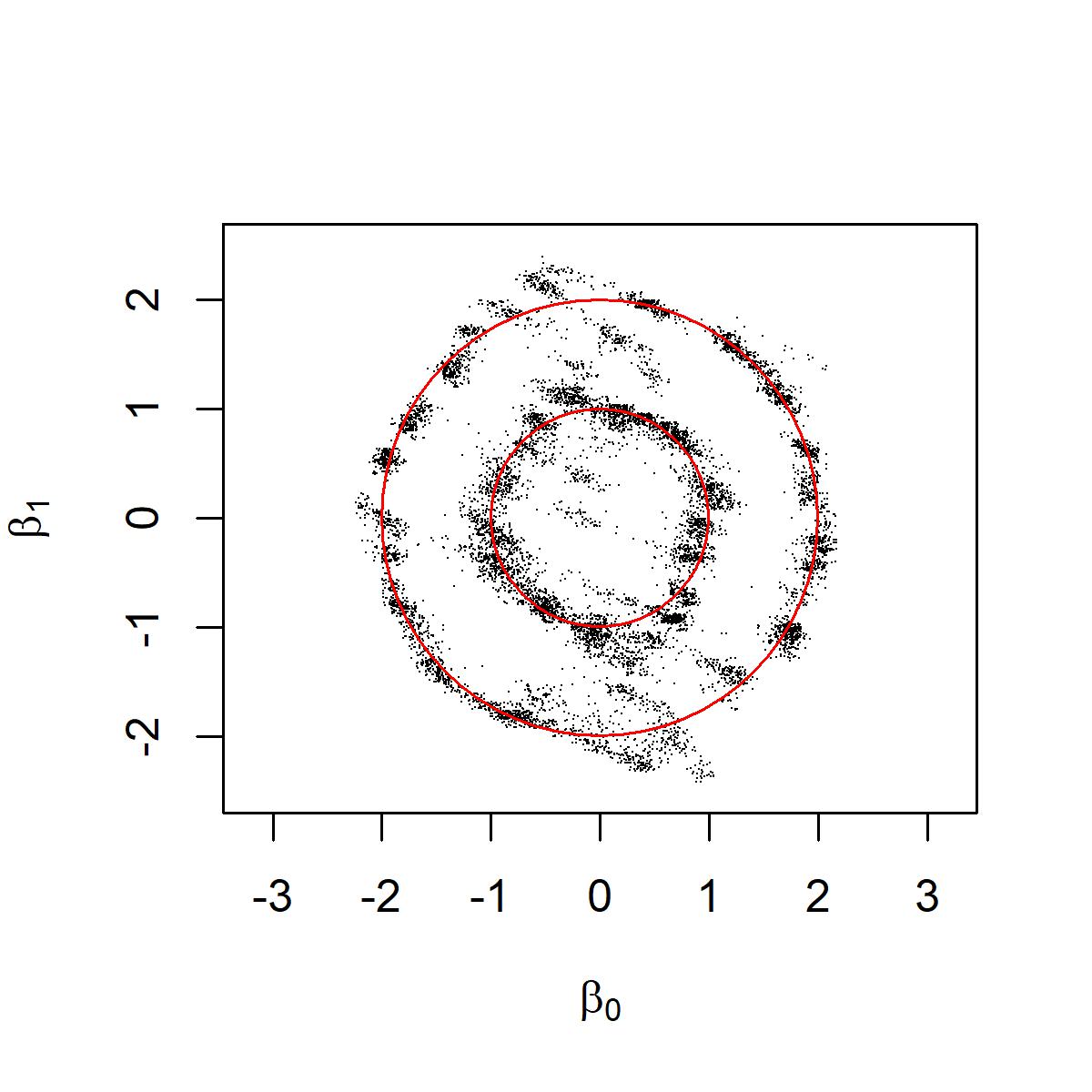}
        \caption{}
        \label{fig:init}
    \end{subfigure}
    \hfill
    \begin{subfigure}[b]{0.45\textwidth}
        \centering
        \includegraphics[width=\textwidth]{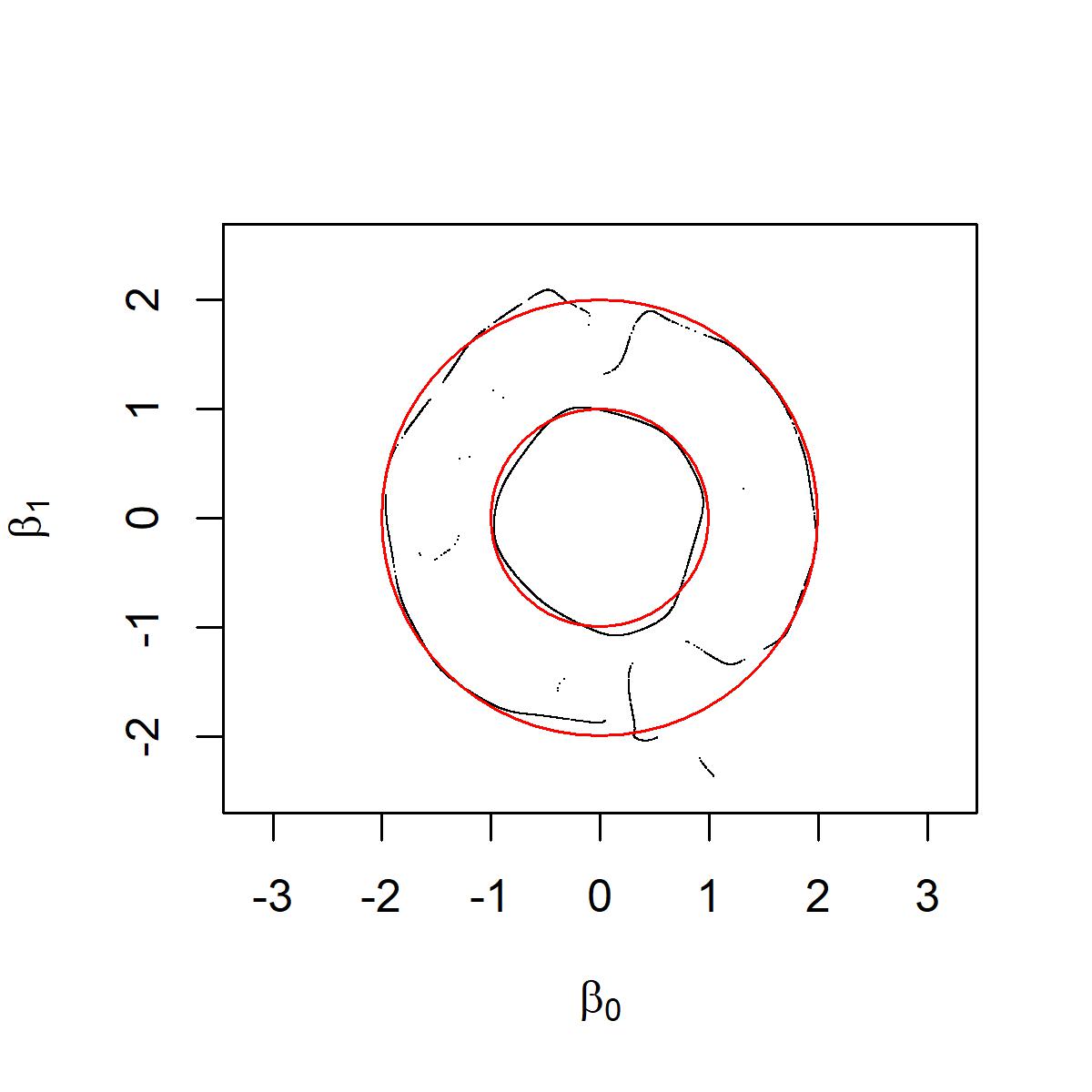}
        \caption{}
        \label{fig:scms}
    \end{subfigure}

    \caption{Comparison of the true mixture distribution and the estimated distributions of $\beta$ for different algorithms in Simulation 2, where the red circles are the support of $G^*$, and point sizes are proportional to weights. 
    (\subref{fig:em}) EM-NPKMLE algorithm: 52 support points, weights ranging from 0.0002 to 0.085. 
     (\subref{fig:cgm}) CGM method: 34 support points, weights ranging from 0.011 to 0.054. (\subref{fig:init}) $10^{4}$ points sampled from the output of the EM-NPMLE algorithm, also used as the input for the SCMS algorithm.
    (\subref{fig:scms}) post-processing result after using the SCMS algorithm, where all points have weight$=10^{-4}$. 
    }
    \label{fig:all_four}
\end{figure}

 As shown in Figure \ref{fig:all_four} for a single random sample, the points produced by the EM-NPKMLE algorithm are roughly evenly distributed near the support of $G^*$, while the support does not seem to be recovered well by the CGM method. The distribution produced by the EM-NPMLE algorithm has a visible concentration near the support of $G^*$, and applying the SCMS algorithm as a post-processing step further extracted the 1-dimensional structure.  

 In Table \ref{W2sim2} we show the results of applying the Wasserstein-2 distance between the probability measures $G^*$ and $\hat{G}$ from the applicable algorithms. Using SCMS algorithm for post-processing has the best performance, followed by the EM-NPMLE algorithm without post-processing. As two algorithms without knowing the prior distribution is continuous or requiring a post-processing step, the EM-NPKMLE is an improvement over the CGM.

\begin{table}[ht!]
\centering
\caption{W-2 Distance between $G^*$ and $\hat G$ for the simulation depicted in Figure \ref{fig:all_four}}
\label{W2sim2}
\begin{tabular}{lc}
\hline
Method & W-2 Distance \\ 
\hline
EM-NPKMLE & 0.328 \\ 
CGM       & 0.394 \\ 
EM-NPMLE  & 0.190 \\ 
EM-NPMLE + SCMS      & 0.157 \\ 
\hline
\end{tabular}
\end{table}
\FloatBarrier
\subsection{Applications}
\label{sec:Applications}

\paragraph{Application:} CO2-GDP

We tested our EM-NPKMLE algorithm on data of 159 countries' CO2 emissions and GDP per capita in 2015 taken from \citet{bolt2020maddison,friedlingstein2022global,owidco2andgreenhousegasemissions}; and \citep{UnitedNations}. This dataset was also analyzed in \citet{jiang2025nonparametric}. Identifying distinct components among the relationship between CO2 emissions and economic output could reveal countries that have high economic output and low CO2 emissions, potentially revealing favorable policies to pursue. Other papers have also looked into this relationship from a mixture of regression approach, such as \citet{huang2012mixture} and \citet{hurn2003estimating}, though these papers use different data sources. 
As in \citet{jiang2025nonparametric}, we used data on CO2 emission in per capita in 10-ton increments and GDP per capita in 10 thousand U.S. dollars. 

\begin{figure}[ht!]
    \centering
\includegraphics[width=0.8\textwidth]{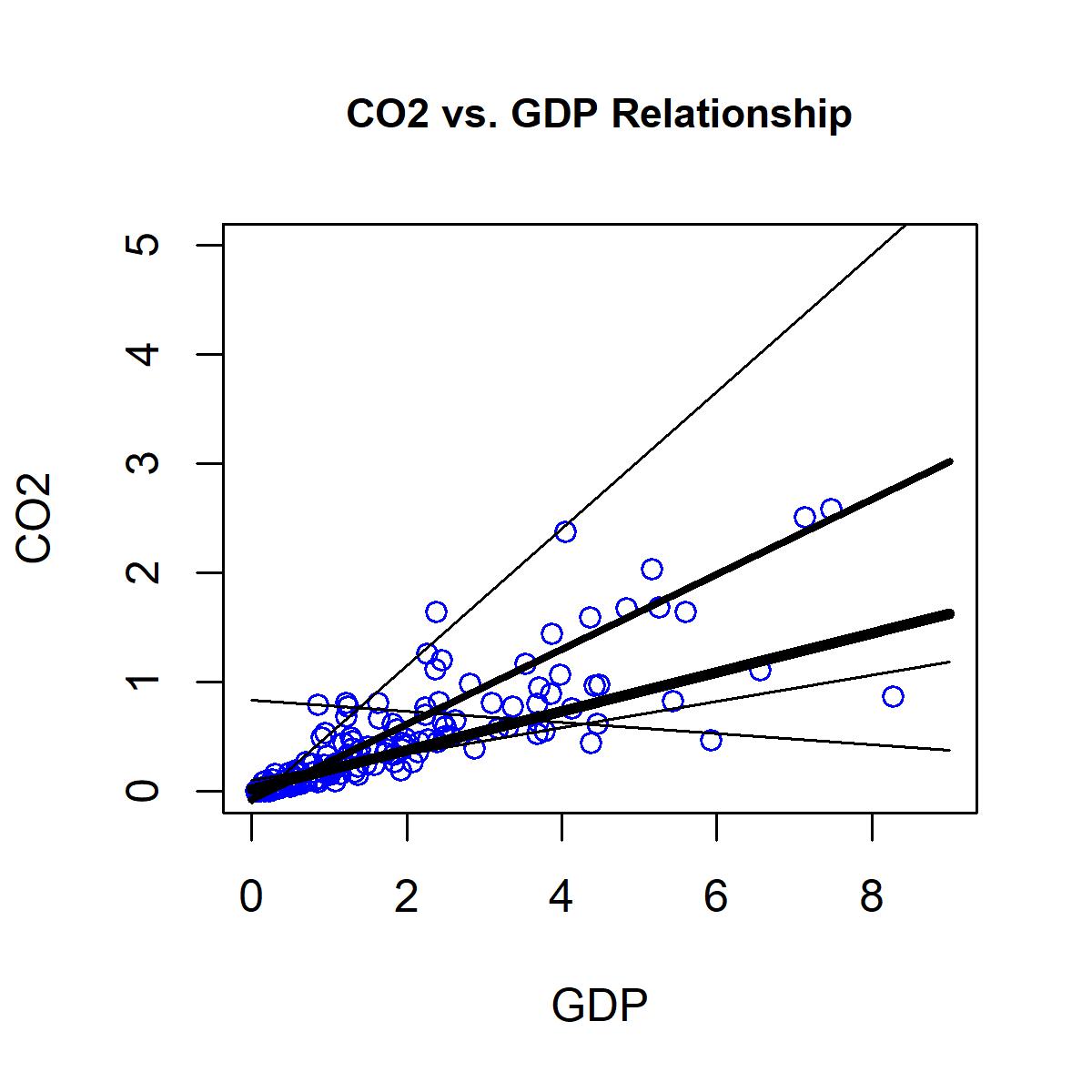} 
\caption{CO2-GDP Relationship: The points represent CO2 emissions in per capita in 10-ton increments and GDP per capita in 10 thousand U.S. dollars for individual countries. The lines represent the estimated component regression lines, whose estimated coefficients can be found in Table \ref{CO2GDPtable}. The thicker lines are the components with larger weights.}
\label{CO2GDPplot}
\end{figure}

We used a 10-fold cross-validation method as described in Section \ref{subsec:simulations} for the CO2-GDP dataset, and used the same bandwidth selection method as described at the beginning of Section \ref{sec:Simulations and Applications}. Our estimate of $\sigma$ from this CV method is 0.160. The results of our algorithm with this estimated $\sigma$ suggest that there are two growth paths with larger estimated prior weights, with possible deviations as evidenced by the other components with smaller weights (see Table \ref{CO2GDPtable} and Figure \ref{CO2GDPplot}). This result differs from \citet{jiang2025nonparametric}, who identified ten components before applying the BIC trimming, and five components after BIC trimming. We note that the two components with largest weights detected from our algorithm ((0.022, 0.179) and (-0.070, 0.343)) are similar to the two components with the largest weights from \citet{jiang2025nonparametric} before and after their BIC trimming method was applied, that is, (0.030, 0.170) and (-0.010, 0.230), with pre-BIC trimming weights of 0.290 and 0.180, and post-BIC trimming weights of 0.270 and 0.280, respectively.

\renewcommand{\arraystretch}{0.7}
\begin{table}[ht!]
\centering
\caption{Results from CO2/GDP Application}
\resizebox{\columnwidth}{!}
{\begin{tabular}{cccccccc}
  \hline
 & 1st Comp & 2nd Comp & 3rd Comp & 4th Comp & 5th Comp \\ 
  \hline
$\beta$ Estimates  & (0.022, 0.179) &  (-0.070, 0.343)  & (-0.101, 0.628)  & (0.105, 0.120) & (0.839, -0.051)   \\ 
Estimated Weights  & 0.484  & 0.358 & 0.088  &  0.057 & 0.013
\\
   \hline
\end{tabular}}
\label{CO2GDPtable}
\end{table}

\paragraph{Application:} Music Tone Perception

\begin{figure}[ht!]
    \centering
\includegraphics[width=0.8\textwidth]{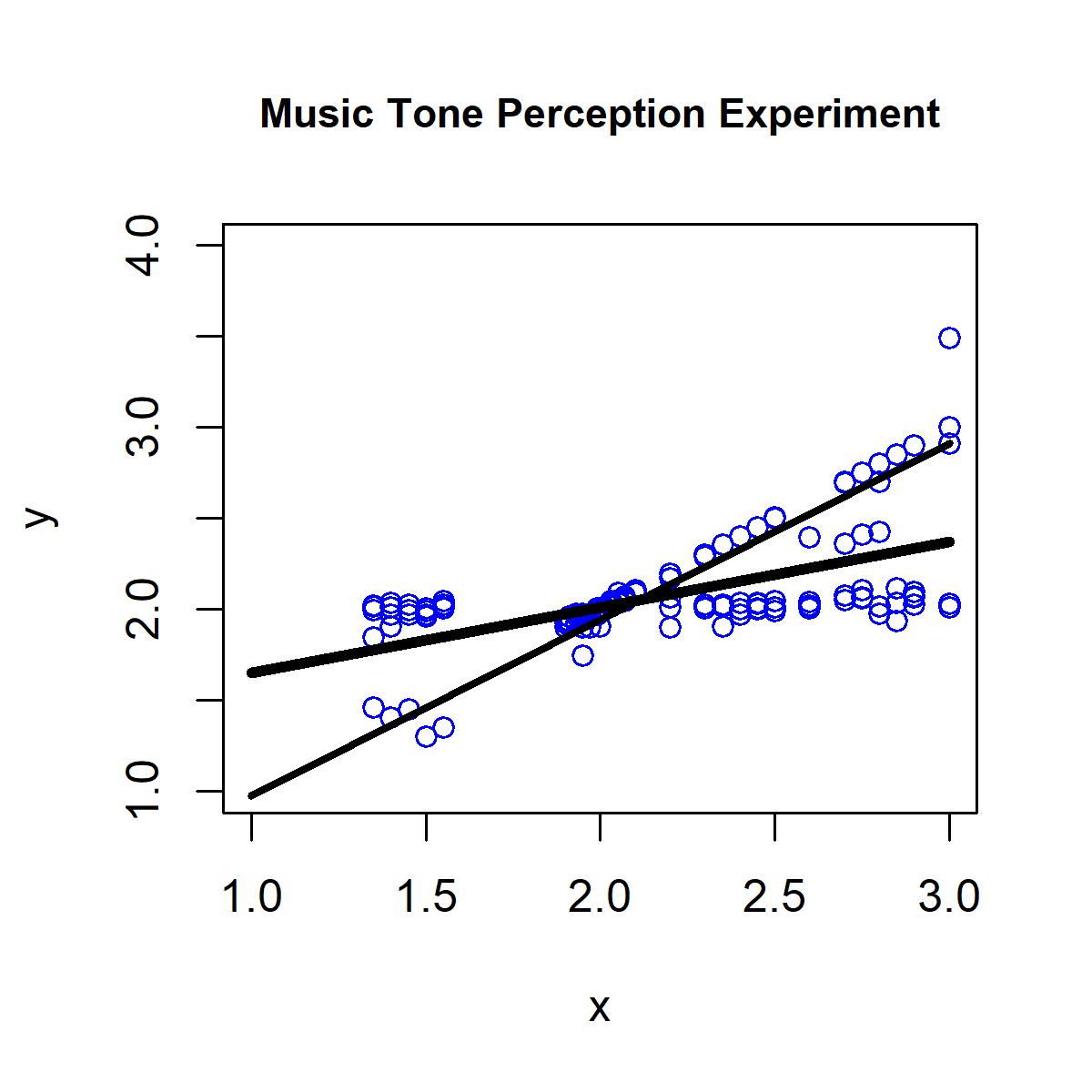}
\caption{Music Tone Perception: The blue points represent a musician's assessed tone ratio ($y$) and the true tone ratio ($x$) pairs. The estimated regression lines (see Table \ref{musictable}) are also depicted for the two components. The thicker line is the component with the larger weight.}
\label{musictone}
\end{figure}

We also tested our EM-NPKMLE algorithm on another dataset analyzed in \citet{jiang2025nonparametric}. The data was initially collected by \citet{cohen1980inharmonic} in an experiment examining how trained musicians perceive tone. See \citet{cohen1980inharmonic} for further information on the experiment. This dataset, which can be found in the {\sf R} packages {\sf mixtools} and {\sf fpc}, contains one musician's tuning that was observed 150 times \citep{mixtools,fpc}. See Figure \ref{musictone}, where the $y$-axis displays the musician's assessed tone ratio and the $x$-axis depicts the true tone ratio \citep{viele2002modeling,yao2015mixtures}. Other studies of this dataset were carried out in \citet{de1989mixtures}; \citet{viele2002modeling}; and \citet{yao2015mixtures}; but in these studies the number of components was pre-specified at two. We applied our EM algorithm that does not require prespecifying the number of components. 

We used the same cross-validation method as in the CO2-GDP application to estimate $\sigma$, and our estimate is 0.210. We also used the same bandwidth selection method as in the CO2-GDP application, that is, maximal smoothing bandwidth. Our results revealed two main components (see Figure \ref{musictone} and Table \ref{musictable}). In terms of the number of components, the results are consistent with the music perception theory when the experiment took place, with one component as $y=x$ and another as $y=2$ \citep{jiang2025nonparametric}. Our results (see Table \ref{musictable}) are comparable to the two largest components computed using the CGM algorithm (both pre- and post BIC trimming) in \citet{jiang2025nonparametric}, that is, (1.898, 0.054), (0, 0.989), with weights 0.540 and 0.260 (pre-BIC trimming), and 0.700 and 0.300 (post-BIC trimming), respectively. However, the CGM method detected six components before the BIC trimming method was applied.

\begin{table}[ht!]
\centering
\caption{Results from Music Tone Application}
\begin{tabular}{cccccccc}
  \hline
 & 1st Comp & 2nd Comp \\ 
  \hline
$\beta$ Estimates  & (0.012, 0.967) & (1.292, 0.361) \\ 
Estimated Weights  & 0.427  & 0.573 &  
\\
   \hline
\end{tabular}
\label{musictable}
\end{table}

\section{Discussion}
\label{sec:conc}

In this paper we first introduce an EM algorithm to approximate NPMLE in the setting of mixture of linear regressions when the prior distribution for the coefficients has a density. We then develop another EM algorithm to approximate NPKMLE that can automatically detect the number of components for a discrete prior distribution, and is able to uncover a continuous prior distribution. We provide some theoretical support for our algorithms as well as demonstrate their performance in simulations and applications. In particular, our simulation results suggest that our EM-NPKMLE algorithm can consistently estimate the true number of components and the $\beta$ coefficients and their associated weights when the prior distribution is discrete. While our EM-NPKMLE algorithm does have useful qualities, it requires the selection of a bandwidth, which can be challenging. We selected the oversmoothing bandwidth based on its ability to avoid the appearance of spurious features that may not belong to the true prior distribution, but further investigation into the bandwidth selection could yield better methods, e.g., adaptive bandwidths for different iterations of the algorithm, or introducing unique bandwidths for each dimension of the data. While we demonstrate the algorithm's ability to uncover both discrete and continuous distributions, investigating its theoretical properties is an area of further research. It is clear that our methods can be extended to the setting of mixture of distributions, and we will investigate this in a separate article.

\appendix

\section{Appendix}

\subsection{Proof of Proposition \ref{proposition_vanilla_alg}}

\begin{proof}
Let the Kullback-Leibler (KL) divergence between distributions $P$ and $Q$ with densities $p$ and $q$ be 
\begin{align}
   D_{\text{KL}}(P\|Q) = \int p(x)\log\frac{p(x)}{q(x)}dx.
\end{align}
We can write
\begin{align}
\label{decompofcomploglike}
&\mathbb{E}_{\bm{\beta}|\bm{x,y},G^{(t)}} L(g|\bm{ x,y,\beta})\\
& = - \sum_{i=1}^n \int f_i^{(t+1)}(\beta)\log \frac{f_i^{(t+1)}(\beta)}{g(\beta)}  d\beta + \sum_{i=1}^n \int \log\left[ \phi_\sigma\left( y_i - x_i^\top \beta\right)  f_i^{(t+1)}(\beta)\right]f_i^{(t+1)}(\beta) d\beta \\
& = - \sum_{i=1}^n D_{\text{KL}}(F_i^{(t+1)}\|G) + \sum_{i=1}^n \int \log\left[ \phi_\sigma\left( y_i - x_i^\top \beta\right)  f_i^{(t+1)}(\beta)\right]f_i^{(t+1)}(\beta) d\beta,
\end{align}
where $F_i^{(t+1)}$ is the distribution associated with $f_i^{(t+1)}$.
Hence, maximizing $\mathbb{E}_{\bm{\beta}|\bm{x,y},G^{(t)}} L(g|\bm{ x,y,\beta})$ over $g$ is equivalent to minimizing $\frac{1}{n} \sum_{i=1}^n D_{\text{KL}}(F_i^{(t+1)}\|G)$ over $G\in\mathcal{G}_{\text{den}}$. Let $g^{(t+1)} = \frac{1}{n} \sum_{i=1}^n f_i^{(t+1)}$. Notice that 
\begin{align}
\label{KLminimizer}
    \frac{1}{n} \sum_{i=1}^n \frac{\delta D_{\text{KL}}(F_i^{(t+1)}\|G)}{\delta g} \bigg|_{g= g^{(t+1)}} = - \frac{1}{n}\sum_{i=1}^n \frac{f_i^{(t+1)}}{g^{(t+1)}} = -1.
\end{align}

\noindent Using Theorem 2 in \citep{nishiyama2020minimization}, we conclude that $g^{(t+1)}$ is the unique minimizer of \\ 
$\frac{1}{n} \sum_{i=1}^n D_{\text{KL}}(F_i^{(t+1)}\|G)$, and hence the unique maximizer of $\mathbb{E}_{\bm{\beta}|\bm{x,y},G^{(t)}} L(g|\bm{ x,y,\beta},G^{(t)})$. 
\end{proof}

\subsection{Proof of Theorem \ref{Qincreasing}}

\begin{proof} 
Using the subgradient inequality for differentiable convex functions, we have 
\[\log[\sum_{\ell=1}^n v(y_\ell)]-\log[\sum_{\ell=1}^n v(x_\ell)]\geq \frac{\sum_{\ell=1}^n v^\prime(x_\ell)(y_\ell-x_\ell)}{\sum_{\ell=1}^n v(x_\ell)}, \quad \forall x_\ell,y_\ell \in [0,\infty).\]
Also, recall that $w=-v^\prime$. Then
\begin{align}
&Q(G_{\bm{\nu^{(r+1)}}};G^{(t)})-Q(G_{\bm{\nu^{(r)}}};G^{(t)})\\
&=\sum_{i=1}^n  \frac{\int_{\mathbb{R}^d}  \log \left[\phi_\sigma\left( y_i - x_i^\top \beta\right) \frac{1}{nh^d} \sum_{\ell=1}^n v_h(\|\beta - \nu_\ell^{(r+1)}\|^2) \right]S_i(\beta,\bm{\beta}^{(t)},\bm{x},\bm{y})d\beta}{\int_{\mathbb{R}^d}  S_i(\beta,\bm{\beta}^{(t)},\bm{x},\bm{y}) d\beta}\\
&\hspace{1cm}-\sum_{i=1}^n  \frac{\int_{\mathbb{R}^d}  \log \left[\phi_\sigma\left(y_i - x_i^\top \beta\right) \frac{1}{nh^d} \sum_{\ell=1}^n v_h(\|\beta-\nu_\ell^{(r)}\|^2) \right]S_i(\beta,\bm{\beta}^{(t)},\bm{x},\bm{y})d\beta}{\int_{\mathbb{R}^d}  S_i(\beta,\bm{\beta}^{(t)},\bm{x},\bm{y}) d\beta}\\
&\geq \sum_{i=1}^n  \frac{\int_{\mathbb{R}^d} \frac{\sum_{\ell=1}^n w_h\left(\|\beta-\nu_\ell^{(r)}\|^2\right)[\|\beta-\nu_\ell^{(r)}\|^2-\|\beta-\nu_\ell^{(r+1)}\|^2]}{\sum_{j=1}^n v_h\left(\|\beta - \nu_j^{(r)}\|^2\right)} S_i(\beta,\bm{\beta}^{(t)},\bm{x},\bm{y})d\beta}{\int_{\mathbb{R}^d}  S_i(\beta,\bm{\beta}^{(t)},\bm{x},\bm{y}) d\beta} \\
&= \sum_{i=1}^n  \frac{\int_{\mathbb{R}^d} \frac{\sum_{\ell=1}^n w_h\left(\|\beta-\nu_\ell^{(r)}\|^2\right)[2\nu_\ell^{(r+1)\top}\beta-2\nu_\ell^{(r)\top}\beta-\|\nu_\ell^{(r+1)}\|^2+\|\nu_\ell^{(r)}\|^2]}{\sum_{j=1}^n v_h\left(\|\beta- \nu_j^{(r+1)}\|^2\right)}S_i(\beta,\bm{\beta}^{(t)},\bm{x},\bm{y})d\beta}{\int_{\mathbb{R}^d}  S_i(\beta,\bm{\beta}^{(t)},\bm{x},\bm{y}) d\beta} \\
&=\sum_{\ell=1}^n2\nu_\ell^{(r+1)\top}\sum_{i=1}^n  \frac{\int_{\mathbb{R}^d} \frac{w_h\left(\|\beta-\nu_\ell^{(r)}\|^2\right)\beta}{\sum_{j=1}^n v_h\left(\|\beta- \nu_j^{(r)}\|^2\right)}S_i(\beta,\bm{\beta}^{(t)},\bm{x},\bm{y})d\beta}{\int_{\mathbb{R}^d}  S_i(\beta,\bm{\beta}^{(t)},\bm{x},\bm{y}) d\beta}\\
&\hspace{1cm}-\sum_{\ell=1}^n2\nu_\ell^{(r)\top}\sum_{i=1}^n  \frac{\int_{\mathbb{R}^d} \frac{w_h\left(\|\beta-\nu_\ell^{(r)}\|^2\right)\beta}{\sum_{j=1}^n v_h\left(\|\beta- \nu_j^{(r)}\|^2\right)}S_i(\beta,\bm{\beta}^{(t)},\bm{x},\bm{y})d\beta}{\int_{\mathbb{R}^d}  
S_i(\beta,\bm{\beta}^{(t)},\bm{x},\bm{y}) d\beta}\\
&\hspace{1cm}-\sum_{\ell=1}^n\|\nu_\ell^{(r+1)}\|^2\sum_{i=1}^n  \frac{\int_{\mathbb{R}^d} \frac{w_h\left(\|\beta-\nu_\ell^{(r)}\|^2\right)}{\sum_{j=1}^n v_h\left(\|\beta- \nu_j^{(r)}\|^2\right)}S_i(\beta,\bm{\beta}^{(t)},\bm{x},\bm{y})d\beta}{\int_{\mathbb{R}^d}  S_i(\beta,\bm{\beta}^{(t)},\bm{x},\bm{y}) d\beta}\\
&\hspace{1cm}+\sum_{\ell=1}^n\|\nu_\ell^{(r)}\|^2\sum_{i=1}^n  \frac{\int_{\mathbb{R}^d} \frac{w_h\left(\|\beta-\nu_\ell^{(r)}\|^2\right)}{\sum_{j=1}^n v_h\left(\|\beta- \nu_j^{(r)}\|^2\right)}S_i(\beta,\bm{\beta}^{(t)},\bm{x},\bm{y})d\beta}{\int_{\mathbb{R}^d}  S_i(\beta,\bm{\beta}^{(t)},\bm{x},\bm{y}) d\beta}\\
&=\sum_{\ell=1}^n [2\nu_\ell^{(r+1)\top}A(\nu_\ell^{(r)};\bm{\beta^{(t)}},\bm{x,y})-2\nu_\ell^{(r)\top}A(\nu_\ell^{(r)};\bm{\beta^{(t)}},\bm{x,y})\\
&\hspace{1cm} -\|\nu_\ell^{(r+1)}\|^2C(\nu_\ell^{(r)},\bm{\beta^{(t)}},\bm{x,y})+\|\nu_\ell^{(r)}\|^2C(\nu_\ell^{(r)},\bm{\beta^{(t)}},\bm{x,y})].
\end{align}

It then follows from \eqref{Aeq},\eqref{Ceq}, and \eqref{gradvec} that 
\begin{align}
&Q(G_{\bm{\nu^{(r+1)}}};G^{(t)})-Q(G_{\bm{\nu^{(r)}}};G^{(t)})\\
&\ge\sum_{\ell=1}^n -\|\nu_\ell^{(r+1)}\|^2C(\nu_\ell^{(r)},\bm{\beta^{(t)}},\bm{x,y})+2\|\nu_\ell^{(r+1)}\|^2C(\nu_\ell^{(r)};\bm{\beta^{(t)}},\bm{x,y})\\
&\hspace{1cm}-2\nu_\ell^{(r)\top}\nu_\ell^{(r+1)}C(\nu_\ell^{(r)};\bm{\beta^{(t)}},\bm{x,y})+\|\nu_\ell^{(r)}\|^2C(\nu_\ell^{(r)},\bm{\beta^{(t)}},\bm{x,y})]\\
&=\sum_{\ell=1}^n C(\nu_\ell^{(r)},\bm{\beta^{(t)}},\bm{x,y})\|\nu_\ell^{(r+1)}-\nu_\ell^{(r)}\|^2 \label{eq:final} \\
&\geq 0,
\end{align}
where the last inequality holds because 
$C(\nu_\ell^{(r)},\bm{\beta^{(t)}},\bm{x,y}) > 0$ by noticing that $w$ is always positive.  

Next we show that $ Q(G_{\bm{\nu}};G^{(t)})$ has an upper bound. Notice that
\begin{align}
\sup_\beta \left|\phi_\sigma\left( y_i - x_i^\top \beta\right) \frac{1}{nh^d} \sum_{\ell=1}^n v_h(\|\beta - \nu_\ell\|^2) \right| \le \frac{\|v\|_\infty}{\sqrt{2\pi}\sigma h^d} =: B, 
\end{align}
where $\|v\|_\infty=\sup_t |v(t)|$. It then follows from \eqref{Q} that
\begin{align}
\sup_{\bm{\nu}} Q(G_{\bm{\nu}};G^{(t)}) \le n \log B.
\end{align}
Hence the sequence $Q(G_{\bm{\nu^{(r)}}};G^{(t)})$ converges.
\end{proof}

\subsection{Proof of Theorem \ref{incomp_monotonic}}
\begin{proof}
For a distribution $G$ on $\mathbb{R}^d$ with density function $g$, let
\begin{align}
P(\bm{\beta}, G)= \sum_{i=1}^{n} \log(f_i(\beta_i\; | \; x_i,y_i, g)),
\end{align}
where $f_i(\beta_i\; | \; x_i,y_i, g)$ is defined in \eqref{fi}. 
We can write
\begin{align*}
P(\bm{\beta}, G)= & \sum_{i=1}^{n} \log\left[\frac{ \phi_\sigma\left( y_i - x_i^\top \beta_i\right) g(\beta_i) }{\int_{\mathbb{R}^d}  \phi_\sigma\left(y_i - x_i^\top \beta\right) g(\beta) d\beta}\right] \\
 = & \sum_{i=1}^{n} \log\left[ \phi_\sigma\left( y_i - x_i^\top \beta_i\right) g(\beta_i)\right] - \sum_{i=1}^{n} \log \left[\int_{\mathbb{R}^d}  \phi_\sigma\left(y_i - x_i^\top \beta \right) g(\beta)d\beta\right].
\end{align*}
Using the definitions in \eqref{incomplete-log} and \eqref{complete-log},
\begin{align*}
L(G)=L(G| \bm{x,y,\beta})  - P(\bm{\beta}, G).
\end{align*}
\noindent Taking the expectation of each side with respect to $\bm{\beta}|\bm{x,y},G^{(t)}$, we have
\begin{align*}
L(G) = \mathbb{E}_{\bm{\beta}|\bm{x,y},G^{(t)}}(L(G)) & = \mathbb{E}_{\bm{\beta}|\bm{x,y},G^{(t)}}(L(G| \bm{x,y,\beta})) - \mathbb{E}_{\bm{\beta}|\bm{x,y},G^{(t)}}(P(\bm{\beta}, G))\\
& =: Q(G;G^{(t)}) - H(G;G^{(t)}).
\end{align*}
Notice that
\begin{align*}
L(G^{(t+1)})- L(G^{(t)}) = \left[ Q(G^{(t+1)};G^{(t)}) - Q(G^{(t)};G^{(t)})\right]  - \left[H(G^{(t+1)};G^{(t)})-H(G^{(t)};G^{(t)}) \right].
\end{align*}
We have already shown in Theorem \ref{Qincreasing} that $Q(G^{(t+1)};G^{(t)}) \geq Q(G^{(t)};G^{(t)})$. 
To show $L(G^{(t+1)})- L(G^{(t)}) \geq 0$, it remains to prove that 
\[H(G^{(t+1)};G^{(t)})-H(G^{(t)};G^{(t)}) \leq 0.\]
This is true because for any arbitrary distribution $G$ with density function $g$,
\begin{align*}
& H(G;G^{(t)})-H(G^{(t)},G^{(t)})\\
=&\mathbb{E}_{\bm{\beta}|\bm{x,y},G^{(t)}} \left[\sum_{i=1}^{n} \log(f_i (\beta_i\; | \; x_i,y_i, g ))\right]-\mathbb{E}_{\bm{\beta}|\bm{x,y},G^{(t)}} \left[\sum_{i=1}^{n} \log(f_i (\beta_i\; | \; x_i,y_i, g^{(t)} ))\right]\\
=&\mathbb{E}_{\bm{\beta}|\bm{x,y},G^{(t)}} \left\{\sum_{i=1}^{n} \log\left[\frac{f_i (\beta_i\; | \; x_i,y_i, g )}{f_i (\beta_i\; | \; x_i,y_i, g^{(t)} )}\right]\right\}\\
= &\sum_{i=1}^{n} \mathbb{E}_{\bm{\beta}|\bm{x,y},G^{(t)}}\left\{\log\left[\frac{f_i (\beta_i\; | \; x_i,y_i, g )}{f_i (\beta_i\; | \; x_i,y_i, g^{(t)} )}\right]\right\}\\
\leq &\sum_{i=1}^{n} \log \left\{\mathbb{E}_{\bm{\beta}|\bm{x,y},G^{(t)}}\left[\frac{f_i (\beta_i\; | \; x_i,y_i, g )}{f_i (\beta_i\; | \; x_i,y_i, g^{(t)} )}\right]\right\}    \noindent\\
&\text { (by Jensen's inequality and the concavity of the log function)}\\
=&\sum_{i=1}^{n}\log\left[\int \frac{f_i (\beta\; | \; x_i,y_i, g )}{f_i (\beta\; | \; x_i,y_i, g^{(t)} )} f_i (\beta\; | \; x_i,y_i, g^{(t)}) d\beta \right]\\
=&\sum_{i=1}^{n}\log\left[\int f_i (\beta\; | \; x_i,y_i, g) d\beta \right]\\
=&0.
\end{align*}
We can also conclude from Theorem 1 in \citep{jiang2025nonparametric} that the incomplete log likelihood $L(G)$ is bounded from above because of the existence of a maximizer $\hat{G}$. By the monotone convergence theorem, $L(G^{(t)})$ converges.
\end{proof}

\section*{Acknowledgments}

This work was partially supported by resources provided by the Office of Research Computing at George Mason University (URL: \url{https://orc.gmu.edu}).

\section*{Data Availability Statement}

The music tone data that supports the findings of this study is openly available in the {\sf R} packages {\sf mixtools} (https://cran.r-project.org/web/packages/mixtools/index.html), {\sf fpc} (https://cran.r-project.org/web//packages/fpc/index.html). The CO2-GDP data can be found at https://github.com/hanshengjiang/npmle\_git. 

\bibliographystyle{apalike} 
\bibliography{bibfile.bib}
\end{document}